\documentclass[12pt]{article}
\usepackage{wire-tap}
\begin{document}
\title{Capacity Region for\\
Quantum Wiretap Coding}

\author{Robert R. Tucci\\
        P.O. Box 226\\
        Bedford,  MA   01730\\
        tucci@ar-tiste.com}

\date{\today}
\maketitle
\vskip2cm
\section*{Abstract}
This paper follows
very closely a famous paper
by Csisz\'{a}r and K\"{o}rner
about classical (non-quantum)
wiretap coding.
Our paper
gives a self-contained
and slightly novel
review of some important
results
of
the paper by Csisz\'{a}r and K\"{o}rner.
Then we present
a generalization of
those results to the quantum realm,
thus giving one of the first
half-decent treatments of
quantum wiretap coding.
Like Csisz\'{a}r and K\"{o}rner,
we too find a capacity region
(i.e., the maximal achievable
region of rates)
characterized in terms of
one-letter informations.
We try to make
our treatment of
quantum wiretap coding
as parallel a possible
to our treatment
of
classical
wiretap coding.
This
parallel treatment
is facilitated
by the use
of CB nets (classical Bayesian networks)
for the classical case and
QB nets (quantum Bayesian networks)
 for the quantum one.

\newpage

\section{Introduction}

For a good textbook on classical (non-quantum)
Shannon
Information Theory (SIT), see, for example,
Ref.\cite{CovTh}
by Cover and Thomas.
For a good
textbook
on quantum SIT, see, for example,
Ref.\cite{Wilde} by Wilde.
Wiretap
coding is an example
of network information theory,
which in turn is part of SIT.
For a good textbook on classical
network information theory,
see, for example,
Ref.\cite{ElG} by El Gamal and Kim.

Wiretap coding
is
a famous topic in classical
SIT.
In classical discussions of wiretap coding,
one considers a situation wherein
 Alice sends a private
message to Bob and a
shared message to both Bob and Eve.
Many workers
have pointed out that an analogous
situation arises
when doing quantum communication.
In that case, the
environment
plays the role of Eve.

The subject
of classical wiretap coding
has an illustrious history.
In Ref.\cite{Wy},
Wyner
found the so called
capacity region
(i.e., the maximal achievable
region of rates) for
a special type of
channel, a ``degraded
wiretap channel".
Wyner characterized his
capacity region using one-letter informations.
In Ref.\cite{CK},
Csisz\'{a}r and K\"{o}rner
generalized
the results of Wyner
to encompass any type
of discrete memoryless
wiretap channel.
They removed the degraded
channel assumption, but
still characterized the
capacity region in terms of one-letter informations.

This paper follows
Ref.\cite{CK}
by Csisz\'{a}r and K\"{o}rner very closely.
Our paper
gives a self-contained
and slightly novel
review of some important
results
of
Ref.\cite{CK}.
Then it presents
a generalization of
those results to the quantum realm,
thus giving one of the first
half-decent treatments of
quantum wiretap coding.
Like Csisz\'{a}r and K\"{o}rner,
we too find a capacity region
characterized in terms of
one-letter informations.

We try to make
our treatment of
quantum wiretap coding
as parallel a possible
to our treatment
of
classical
wiretap coding.
This
parallel treatment
is facilitated
by the use
of CB nets (classical Bayesian networks)
for the classical case and
QB nets (quantum Bayesian networks)
 for the quantum one.

The capacity regions found by Wyner,
then Csisz\'{a}r/K\"{o}rner, and now us,
are 3 dimensional. They are
a closed, convex
set of points $(R_\rve,R_\rvs,R_\rvt)$.
Here $R_\rvs$
is the rate of the secret, private
message, $R_\rvt$
is the rate of the tapped, shared message,
and
$R_\rve$ is the so called
equivocation rate
which measures how ignorant
Eve is about the secret message $\rvs$.

Many authors
have previously discussed
the topic of
sending classical
and quantum messages
through a quantum channel,
in some cases
with ``entanglement assistance".
See Ref.\cite{Wilde}
for a masterful review
with ample references.
However,
few authors have
specifically
discussed
 quantum wiretap
coding.
No previous paper,
as far as I know,
gives a treatment
of quantum wiretap coding
that includes
proofs of
achievability
and optimization
for
a rate region
characterized by
one-letter informations.

This paper is written assuming that
the reader has first read
3 previous papers by the same
author.
\begin{itemize}
\item
Ref.\cite{Tuc-mixology}
is an introduction
to quantum Bayesian networks for mixed states.
\item
Ref.\cite{Tuc-redoing-classics}
redoes the classics (the basic
theorems of
classical SIT)
using p-type integration techniques
and CB nets.
\item
Refs.\cite{Tuc-inequalities}
discusses well-known
inequalities of classical
and quantum SIT from
a Bayesian networks perspective.
\end{itemize}

\section{Preliminaries and Notation}

Reading Refs.\cite{Tuc-mixology},
\cite{Tuc-redoing-classics}
and \cite{Tuc-inequalities}
is a
prerequisite to reading this paper.
This
section
will introduce
only notation
which  hasn't
been defined already in
those 3 papers.

We will use the
symbol $(y\leftrightarrow z)$
after an equation
to indicate that the
same equation is also valid
with $y$ and $z$ swapped.

Suppose
$n$ is any positive integer.
Let $\rvx^n = (\rvx_1, \rvx_2, \ldots, \rvx_n)$
be the random variable that takes on values
$x^n = (x_1, x_2, \ldots, x_n)\in S_\rvx^n$.
For any integer $j$ such that $1<j<n$, let
$x_{<j} = (x_1, x_2, \ldots, x_{j-1})$,
and
$x_{>j} = (x_{j+1}, x_2, \ldots, x_n)$.

Given $\{P(x)\}_{\forall x}\in pd(S_\rvx)$,
we used in previous papers
the expected value operator:

\beq
E_x = \sum_x P(x)
\;.
\eeq
Given a probability amplitude
$A(x)$ such that $\sum_x |A(x)|^2=1$,
we will now use the
following 2 operators which are sort of
``square roots" of $E_x$:

\beq
\cale_x = \sum_x A(x)
\;,\;\;
\cale^*_x = \sum_x A^*(x)
\;.
\eeq

For both
CB nets and QB nets,
define a {\bf source (or root)
node} as a node that
has only outgoing arrows, no incoming ones.
Define a {\bf sink (or leaf) node}
as one that has only incoming arrows,
no outgoing ones.
Sometimes, to
simplify a graph,
rather than drawing a
root node $\rvr$ and its
outgoing arrows
explicitly,
we will just put
the line
$\rvr\rarrow
\rvx_1,\rvx_2,\ldots,\rvx_n$
beneath the graph,
where
$\rvx_1,\rvx_2,\ldots,\rvx_n$
are the children nodes of $\rvr$.
Likewise, sometimes, rather than drawing
a leaf node $\rvl$ and its
incoming arrows
explicitly,
we will just put
the line
$\rvl\larrow
\rvx_1,\rvx_2,\ldots,\rvx_n$
beneath the graph,
where
$\rvx_1,\rvx_2,\ldots,\rvx_n$
are the parent nodes of $\rvl$.

For QB nets, a {\bf reservoir}
is defined as a {\it sink or source} node
that is traced over.
A reservoir that is a sink
can always be
traded for one that is a source
or vice versa. Indeed, if

\beq
\rho_\rvx
=
\sum_r
\sandb{
\begin{array}{r}
\sum_x
A(x|r)\ket{x}
\\
A(r)\;\;\;\;
\end{array}
}
\sandb{\hc}
=
\tr_\rvr
\sandb{
\entrymodifiers={++[o][F-]}
\xymatrix{
\rvx&\rvr\ar[l]
}}
\sandb{\hc}
\;,
\label{eq-res-past}
\eeq
and we
define amplitudes
$A(r|x)$ and $A(x)$ so that
\beq
A(x|r)A(r)=A(r|x)A(x)
\;
\eeq
for all $r,x$,
then we can also write

\beq
\rho_\rvx
=
\sum_r
\sandb{
\begin{array}{r}
\sum_x
A(x)\ket{x}
\\
A(r|x)\;\;\;\;
\end{array}
}
\sandb{\hc}
=
\tr_\rvr
\sandb{
\entrymodifiers={++[o][F-]}
\xymatrix{
\rvr&\rvx\ar[l]
}}
\sandb{\hc}
\;.
\eeq

Eq.(\ref{eq-res-past}),
call it the ``reservoir as past history" version,
is more directly related
to the eigenvalue decomposition of
the density matrix $\rho_\rvx$.
Indeed, suppose $\rho_\rvx$
has eigenvalues $\lam_x$ and
eigenvectors $\ket{\lam_x}$
for all $x\in S_\rvx$.
Then one can set $S_\rvr=S_\rvx$,
$A(x|r)=\av{x|\lam_r}$
and $A(r)=\sqrt{\lam_r}$.
This makes $A(x|r)$ an isometry.

Note that reservoirs can be merged.
For example,

\beq
\sum_{r_1,r_2}
\sandb{
\begin{array}{r}
\sum_{a,b,c}A(a)\ket{a}
\\
A(r_1|a,c)A(b|a)\ket{b}
\\
A(r_2|c)A(c|b)\ket{c}
\end{array}
}
\sandb{\hc}
=
\sum_{r}
\sandb{
\begin{array}{r}
\sum_{a,b,c}A(a)\ket{a}
\\
A(b|a)\ket{b}
\\
A(r|a,c)A(c|b)\ket{c}
\end{array}
}
\sandb{\hc}
\;,
\eeq
where $r=(r_1,r_2)$.
In diagrammatic
language,

\beq
\tr_{\rvr_1,\rvr_2}
\sandb{
\entrymodifiers={++[o][F-]}
\xymatrix{
\rvr_1&*{}&\rva\ar[ll]\ar[d]
\\
\rvr_2&\rvc\ar[l]\ar[lu]&\rvb\ar[l]
}
}
\sandb{\hc}
=
\tr_\rvr
\sandb{
\entrymodifiers={++[o][F-]}
\xymatrix{
\rvr&*{}&\rva\ar[ll]\ar[d]
\\
*{}&\rvc\ar[lu]&\rvb\ar[l]
}
}
\sandb{\hc}
\;.
\eeq
This example involved merging
2 sink reservoirs. Two source reservoirs
can likewise be merged.

We will often denote the eigenvalues
and eigenvectors of
a matrix $\{M_{x,y}\}_{\forall x,y\in S_\rvx}$
by $\lam_x(M)$ and $\ket{\lam_x(M)}$
for all $x\in S_\rvx$.
Note that if $\rho_\rvx \in dm(\calh_\rvx)$,
then
$\{\lam_x(\rho_\rvx)\}_{\forall x}\in pd(S_\rvx)$.
 Also if $\rho_{\rvx,\rvy}\in
 dm(\calh_{\rvx,\rvy})$, then
$\{\lam_{x,y}(\rho_{\rvx,\rvy})\}_{\forall x,y}
\in pd(S_{\rvx,\rvy})$.
Let $\lam_{x,y,e}=
\lam_{x,y,e}(\rho_{\rvx,\rvy,\rve})$ for
some fixed
density matrix $\rho_{\rvx,\rvy,\rve}$.
Since $\lam_{x,y,e}$
behaves like a classical probability distribution,
it is
convenient to define $\lam_{x,y}=\sum_e \lam_{x,y,e}$,
$\lam_{y|x} = \frac{\lam_{x,y}}{\lam_x}$,
$\lam_{y:x} = \frac{\lam_{x,y}}{\lam_x\lam_y}$,
etc.

Consider a
density matrix $\rho_{\rvx,\rvy}$.
Just like we refer to
$\rho_\rvy = \tr_\rvx\rho_{\rvx,\rvy}$
as a marginal (or partial, or reduced) density matrix of
$\rho_{\rvx,\rvy}$,
we will refer
to
$\lam_y(\rho_{\rvx,\rvy}) =
\sum_x \lam_{x,y}(\rho_{\rvx,\rvy})$
as a marginal (or partial) eigenvalue of
$\rho_{\rvx,\rvy}$. It is important to realize that

\beq
\lam_y(\rho_{\rvx,\rvy})
\neq
\lam_y(\rho_{\rvy})
\;.
\label{eq-eve-not-equal}
\eeq
That is, in general, a marginal eigenvalue
of
$\rho_{\rvx,\rvy}$
is not equal to the
eigenvalue of a marginal
density matrix of $\rho_{\rvx,\rvy}$.
The provenance of the
eigenvalue matters.
As with famous paintings, one can't trace
away part of that provenance with impunity.
For a diagonal density matrix
$\rho_{\rvx,\rvy}$ (i.e.,
the classical case),
Eq.(\ref{eq-eve-not-equal}) becomes an equality.

\section{Conditioning on Two Parents}
The
following inequality
will prove useful later on.

\begin{claim}\label{claim-two-parents}
If $\rho_{\rva,\rvx,\rvy}
\in dm(\calh_{\rva,\rvx,\rvy})$,
then

\beq
0\leq S(\rva|\rvx) + S(\rva|\rvy)
\;.
\label{eq-two-parents}
\eeq
\end{claim}
\proof
Suppose
$\rho_{\rva,\rvx,\rvy,\rvr}
\in dm(\calh_{\rva,\rvx,\rvy,\rvr})$
is a pure state with partial trace
$\rho_{\rva,\rvx,\rvy}$.
Because of CMI $\geq 0$, we must have

\beq
S(\rva|\rvx,\rvr)
\leq
S(\rva|\rvx)
\;.
\eeq
Hence

\beq
S(\rva,\rvx,\rvr)-S(\rvx,\rvr)
\leq
S(\rva|\rvx)
\;.
\label{eq-two-parents-pre-schmidt}
\eeq
But
since $\rho_{\rva,\rvx,\rvy,\rvr}$
is pure, Eq.(\ref{eq-two-parents-pre-schmidt})
implies that

\beq
S(\rvy)-S(\rvy,\rva)
\leq
S(\rva|\rvx)
\;.
\eeq
Thus

\beq
-S(\rva|\rvy)
\leq
S(\rva|\rvx)
\;.
\eeq
\qed

I like to interpret inequality
Eq.(\ref{eq-two-parents})
as saying that
conditioning on two parents
yields a net positive entropy.

\section{Chain Rule Extravaganza}

The
theories
for classical and quantum wiretap
coding
both
rely on some identities that are direct
consequences (one might
even call then souped-up chain rules) of the
the chain rules of classical and quantum
SIT.
This section will
derive those
souped-up chain rules.

\emph{All identities in this
 section will be
stated for
classical entropies
$H$, but they
also hold
in the quantum case
if we simply replace all
$H$'s by $S$'s.}

When trying to
find the counterpart in quantum SIT
of a result
in classical SIT, it
is often helpful to
keep in mind that
classical and quantum
SIT
satisfy identical (except that
$H$'s are swapped by $S$'s)
{\it identities} if those
identities follow
from the chain rule.
On the other
hand, they sometimes
satisfy different
 {\it inequalities}.
 (For instance,
 $H(\rva|\rvb)$
 is non-negative
 but $S(\rva|\rvb)$
 can be negative.)

Classically, the chain rule
for probabilities
can be stated as

\beq
P(x^n) = \prod_{j=1}^n
P(x_j|x_{>j})=
\prod_{j=1}^n
P(x_j|x_{<j})
\;
\label{eq-chain-rule-probs-usual}
\eeq
for all $x^n\in S_{\rvx^n}$,
for $n$ random variables $\rvx^n$
that are not necessarily i.i.d..
We see that there are two versions
of this chain rule: one
puts conditions on the past and the other
on the future.

An alternative notation that
I find more symmetrical
is as follows. Eschew the
vertical bar that indicates
that conditions will follow. Instead,
put a superscript of
1 on a random variable that
is being conditioned on,
and put a superscript of 0
on a random variable that
is absent (i.e., it's just
serving as a placeholder).
In this ``binary" notation,
Eq.(\ref{eq-chain-rule-probs-usual})
becomes

\beq
P(x^n) = \prod_{j=1}^n
P(x^0_{<j},x_j,x^1_{>j})=
\prod_{j=1}^n
P(x^1_{<j},x_j,x^0_{>j})
\;.
\eeq
Note that in this binary notation,
one can go from
one version of the chain rule
to the other simply by swapping 0 and
1 superscripts.

The chain rule for
probabilities
immediately implies
a chain rule for
plain entropies
and one
for conditional entropies.
For plain entropies, one
gets the following chain rule

\beq
H(\rvx^n) = \sum_{j=1}^n
H(\rvx_j|\rvx_{>j})=
\sum_{j=1}^n
H(\rvx_j|\rvx_{<j})
\;,
\eeq
or, expressed in binary notation,

\beq
H(\rvx^n) = \sum_{j=1}^n
H(\rvx^0_{<j},\rvx_j,\rvx^1_{>j})=
\sum_{j=1}^n
H(\rvx^1_{<j},\rvx_j,\rvx^0_{>j})
\;.
\eeq
For conditional entropies, on
gets the following chain rule

\beq
H(\rvx:\rvy^n)=
\sum_{j=1}^n
H(\rvx:\rvy_j|\rvy_{>j})
=
\sum_{j=1}^n
H(\rvx:\rvy_j|\rvy_{<j})
\;,
\eeq
or, expressed in binary notation,

\beq
H(\rvx:\rvy^n)=
\sum_{j=1}^n
H(\rvx:\rvy_{<j}^0,\rvy_j,\rvy_{>j}^1)
=
\sum_{j=1}^n
H(\rvx:\rvy_{<j}^1,\rvy_j,\rvy_{>j}^0)
\;.
\eeq

\begin{claim}\label{claim-swap-zero-one}
\beq
H(\rvx:\rvb)-H(\rvx:\rva)=
H(\rvx:\rvb|\rva)-H(\rvx:\rva|\rvb)
\;,
\eeq
or, equivalently,

\beq
H(\rvx:\rva^0,\rvb)-H(\rvx:\rva,\rvb^0)=
H(\rvx:\rva^1,\rvb)-H(\rvx:\rva,\rvb^1)
\;.
\eeq
\end{claim}
\proof

\beqa
H(\rvx: \rva,\rvb)
&=&
H(\rvx: \rvb) + H(\rvx: \rva|\rvb)
\\&=&
H(\rvx: \rvb|\rva) + H(\rvx: \rva)
\;.
\eeqa
\qed

Define
\beq
\Sigma_{<\rv{\lam}}
=
\sum_{j=1}^n H(\rvy_j:\rvz_{>j}|\rvy_{<j},\rv{\lam})
=
\sum_{j=1}^n H(\rvy_j:\rvy_{<j}^1,\rvz_{>j},\rv{\lam}^1)
\;,
\eeq
and

\beq
\Sigma_{>\rv{\lam}}
=
\sum_{j=1}^n H(\rvz_j:\rvy_{<j}|\rvz_{>j},\rv{\lam})
=
\sum_{j=1}^n H(\rvz_j:\rvy_{<j},\rvz_{>j}^1,\rv{\lam}^1)
\;.
\eeq

\begin{claim}\label{claim-lt-is-mt}
\beq
\Sigma_{<\rv{\lam}} =
\Sigma_{>\rv{\lam}}
\;.
\eeq
\end{claim}
\proof
\beq
\Sigma_{<\rv{\lam}} =
\sum_{j=1}^n
\sum_{k=j+1}^n
H(\rvy_j:\rvy^1_{<j},\rvz_k,\rvz^1_{>k},\rv{\lam}^1)
\;.
\eeq

\beqa
\Sigma_{>\rv{\lam}} &=&
\sum_{j=1}^n
\sum_{k=1}^{j-1}
H(\rvz_j:\rvy^1_{<k},\rvy_k,\rvz^1_{>j},\rv{\lam}^1)
\\
&=&
\sum_{j=1}^n
\sum_{k=1}^{j-1}
H(\rvy_k:\rvy^1_{<k},\rvz_j,\rvz^1_{>j},\rv{\lam}^1)
\\
&=&
\sum_{k=1}^n
\sum_{j=1}^{k-1}
H(\rvy_j:\rvy^1_{<j},\rvz_k,\rvz^1_{>k},\rv{\lam}^1)
\;.
\eeqa
Finally, note that
$\sum_{j=1}^n
\sum_{k=j+1}^n$ =
$\sum_{k=1}^n
\sum_{j=1}^{k-1}$
when acting on any function of $(j,k)$.
\qed

Henceforth, we will use the shorthand

\beq
\rv{\alpha}_j = (\rvy_{<j},\rvz_{>j})
\;.
\eeq

\begin{claim}\label{claim-yn-colon-t}
\beq
H(\rvy^n:\rvt)
=\sum_j
H(\rvy_j:\rv{\alpha}_j,\rvt)
-
\sum_j H(\rvy_j:\rvy_{<j})
-\Sigma_{<\rvt}
\;
\eeq

\beq
H(\rvz^n:\rvt)
=\sum_j
H(\rvz_j:\rv{\alpha}_j,\rvt)
-
\sum_j H(\rvz_j:\rvz_{>j})
-\Sigma_{>\rvt}
\;
\eeq
\end{claim}
\proof
Just apply the appropriate chain rule to
the
the first $H$
on left hand side
and the first $H$ on the right hand side
of both equations.
\qed

\begin{claim}\label{claim-two-parts-ck}
\beq
H(\rvy^n:\rvs|\rvt)=
\sum_j
H(\rvy_j:\rvs|\rv{\alpha}_j,\rvt)
+\Sigma_{<\rvt}-\Sigma_{<\rvs,\rvt}
\;
\eeq

\beq
H(\rvz^n:\rvs|\rvt)
=
\sum_j
H(\rvz_j:\rvs|\rv{\alpha}_j,\rvt)
+\Sigma_{>\rvt}-\Sigma_{>\rvs,\rvt}
\;
\eeq
\end{claim}
\proof

\beqa
H(\rvy^n:\rvs|\rvt)
&=&
\sum_j H(\rvy_j:\rvy_{<j}^1,\rvz_{>j}^0,\rvs,\rvt^1)
\\
&=&
\sum_j\left\{
\begin{array}{r}
H(\rvy_j:\rvy_{<j}^1,\rvz_{>j},\rvs^0,\rvt^1)\\
+H(\rvy_j:\rvy_{<j}^1,\rvz_{>j}^1,\rvs,\rvt^1)\\
-H(\rvy_j:\rvy_{<j}^1,\rvz_{>j},\rvs^1,\rvt^1)
\end{array}
\right\}
\\
&=&
\Sigma_{<\rvt}
+\sum_j
H(\rvy_j:\rvs|\rv{\alpha}_j,\rvt)
-\Sigma_{<\rvs,\rvt}
\;
\eeqa
\qed

\begin{claim}\label{claim-ck}
(Csisz\'{a}r \& K\"{o}rner's Extreme Chain Ruling Identity)

\beq
H(\rvy^n:\rvs|\rvt)-
H(\rvz^n:\rvs|\rvt)
=
\sum_j
H(\rvy_j:\rvs|\rv{\alpha}_j,\rvt)-
\sum_j
H(\rvz_j:\rvs|\rv{\alpha}_j,\rvt)
\;
\eeq
\end{claim}
\proof
Follows immediately from
Claims \ref{claim-lt-is-mt} and
\ref{claim-two-parts-ck}
\qed

Define
\beq
\begin{array}{l}
n\delta_y = H(\rvs,\rvt|\rvy^n),
\\
n\delta_{ty} = H(\rvt|\rvy^n),
\\
n\delta_{tz} = H(\rvt|\rvz^n)
\end{array}
\;.
\eeq

\begin{claim}\label{claim-enter-deltas}
\beq
H(\rvs|\rvz^n) =
H(\rvy^n : \rvs |\rvt)
-
H(\rvz^n : \rvs |\rvt)
-H(\rvt|\rvz^n,\rvs)
+ n(\delta_{tz} + \delta_y-\delta_{ty})
\;
\eeq

\beq
H(\rvs|\rvt) =
H(\rvy^n : \rvs |\rvt)
+ n(\delta_y-\delta_{ty})
\;
\eeq

\beq
H(\rvt) =
H(\rvy^n : \rvt)
+ n\delta_{ty}
\;
\eeq

\beq
H(t)=H(\rvz^n : \rvt)
+ n\delta_{tz}
\;
\eeq
\end{claim}
\proof

\beqa
\lefteqn{H(\rvs|\rvz^n)}\nonumber
\\
 &=&
H(\rvs)-H(\rvs : \rvz^n)
\\
&=&
H(\rvs)
+
\left[\begin{array}{r}
H(\rvs:\rvt,\rvz^{n0},\rvy^{n0})
\\
-H(\rvs:\rvt^0,\rvz^{n},\rvy^{n0})
\end{array}\right]
-H(\rvs:\rvt)
\\
&=&
H(\rvs)+
\left[\begin{array}{r}
H(\rvs:\rvt,\rvz^{n1},\rvy^{n0})
\\
-H(\rvs:\rvt^1,\rvz^{n},\rvy^{n0})
\end{array}\right]
-
\left[\begin{array}{r}
H(\rvs:\rvt,\rvy^{n})
\\
-H(\rvs:\rvy^{n}|\rvt)
\end{array}\right]
\\
&=&
H(\rvs)+
H(\rvs:\rvt|\rvz^n)
-H(\rvs:\rvz^n|\rvt)
-H(\rvs:\rvt,\rvy^n)
+H(\rvs:\rvy^n|\rvt)
\\
&=&
\cancel{H(\rvs)}
+
\left[\begin{array}{r}
n\delta_{tz}
\\
\scriptstyle -H(\rvt|\rvz^n,\rvs)
\end{array}\right]
-H(\rvs:\rvz^n|\rvt)
-
\left[\begin{array}{r}
\cancel{H(\rvs)}
\\
-n\delta_y
\\+n\delta_{ty}
\end{array}\right]
+H(\rvs:\rvy^n|\rvt)
\;
\eeqa

\beqa
H(\rvs|\rvt) &=&
H(\rvs:\rvy^n|\rvt)+H(\rvs|\rvy^n,\rvt)
\\
&=&
H(\rvs:\rvy^n|\rvt)+n(\delta_y-\delta_{ty})
\;
\eeqa

\beqa
H(\rvt)
&=&
H(\rvt:\rvy^n) + H(\rvt|\rvy^n)
\\
&=&
H(\rvt:\rvy^n) + n\delta_{ty}
\;
\eeqa
\qed

\begin{claim}\label{claim-opt-identity}
\beq
H(\rvs|\rvz^n)=
\sum_j
H(\rvy_j:\rvs|\rv{\alpha}_j,\rvt)
-
\sum_j
H(\rvz_j:\rvs|\rv{\alpha}_j,\rvt)
-H(\rvt|\rvz^n,\rvs)
+n(\delta_{tz} + \delta_y-\delta_{ty})
\;\label{eq-opt-identity-re}
\eeq

\beq
H(\rvs|\rvt)=
\sum_j
H(\rvy_j:\rvs|\rv{\alpha}_j,\rvt)
+\Sigma_{<\rvt}-\Sigma_{<\rvs,\rvt} + n(\delta_y-\delta_{ty})
\;
\label{eq-opt-identity-rs}
\eeq

\beq
H(\rvt)=
\sum_j
H(\rvy_j:\rv{\alpha}_j,\rvt)
-
\sum_j H(\rvy_j:\rvy_{<j})
-\Sigma_{<\rvt}
+ n\delta_{ty}
\;
\label{eq-opt-identity-rty}
\eeq

\beq
H(\rvt)=
\sum_j
H(\rvz_j:\rv{\alpha}_j,\rvt)
-
\sum_j H(\rvz_j:\rvz_{>j})
-\Sigma_{<\rvt}
+n\delta_{tz}
\;
\label{eq-opt-identity-rtz}
\eeq
\end{claim}
\proof
Combine Claims
\ref{claim-lt-is-mt} to
\ref{claim-enter-deltas}.
\qed

\section{Revisiting Classical Channel Coding}

We will assume that the
reader
of this paper
has first read Ref.\cite{Tuc-redoing-classics}.
Ref.\cite{Tuc-redoing-classics}
provides a derivation of
the capacity for
classical channel coding,
a derivation
based on
p-type integration
techniques.
The same p-type integration techniques
will be used
in this paper to
derive capacities for
classical and quantum wiretap coding.

In this section, we will
mention just a few highlights
of the theory
of classical channel coding
as given in Ref.\cite{Tuc-redoing-classics}.
We will state
those highlights
in a form that is
slightly more general
than the form
presented in Ref.\cite{Tuc-redoing-classics}.
The more general form clarifies
the connections
between
channel coding and
wiretap
coding.

Channel coding
is based on the following CB net:

\beq
\begin{array}{c}
\entrymodifiers={++[o][F-]}
\xymatrix{
\what{\rvm}&\rvy^n\ar[l]&\rvx^n\ar[l]&\rvm\ar[l]\\
*{}&\rv{\calc}\ar[ul]\ar[ur]&*{}&*{}
}
\end{array}
\;.
\label{eq-ch-net}
\eeq

The probability distribution
$P_{\rvy|\rvx}$,
called the channel, is
given and fixed. Define\footnote{gen=``general"}

\beq
\calp_{gen} =
\{ P_{\rvx,\rvy}\in pd(S_{\rvx,\rvy}):
P_{\rvy|\rvx}\mbox{ fixed }\}
\;.
\eeq

For any
$P\in pd(S_{\rvx,\rvy})$,
we can define
a convex hull of
$R$'s by

\beq
\calr(P)=
\{R_\rvm: R_\rvm\leq H_P(\rvy:\rvx)\}
\;.
\eeq
It's also useful to consider, for any
$\calp\subset pd(S_{\rvx,\rvy})$, the
set

\beq
\calr(\calp)=
\bigcup_{P\in \calp} \calr(P)
\;.
\eeq

The following Lemma
will be used later on.

\begin{claim}\label{claim-cmi-leq-mi}
Suppose $\rva(\rvj)$
and $\rvb(\rvj)$ are two
random variables that
are functions of a random variable
$\rvj$. Furthermore, suppose
$\rvb(\rvj)=(\rv{\beta},\rvj)$
for some random variable $\rv{\beta}$.
Then\footnote{Note that
the inequality
in Eq.(\ref{eq-knows-j-ineq})
is different
from $H(\rva|\rvj)\leq H(\rva)$,
which is just MI$\geq 0$.}

\beq
H(\rva(\rvj):\rvb(\rvj)|\rvj)
\leq
H(\rva(\rvj):\rvb(\rvj))
\;.
\label{eq-knows-j-ineq}
\eeq
\end{claim}
\proof

\beqa
H(\rva(\rvj):\rvb(\rvj))
&\stackrel{(i)}{=}&
H(\rva(\rvj):\rvb(\rvj),\rvj)
\\
&\stackrel{(ii)}{\geq}&
H(\rva(\rvj):\rvb(\rvj)|\rvj)
\;.
\eeqa
(i) follows by cloning $\rvj$, which
is a component
of $\rvb(\rvj)$.
(ii) follows from the chain rule.
\qed

Later on, we will
use random variables $\rva_j$
that are labeled by an index $j=1,2,\ldots,n$.
The index $j$ can be promoted
to a random variable
that is uniformly distributed.
$\rvj$ is
often called a ``time sharing variable".
We can then use $E_j = \frac{1}{n}\sum_j$.
$\rva_\rvj$
can be thought of as a special case of
 $\rva(\rvj)$
 when $\rvj$
 is uniformly distributed.

\begin{claim}\label{claim-ch-ach}
Achievability:
$\forall R_\rvm$, if $R_\rvm\in \calr(\calp_{gen})$, then
$\exists$
an encoding and a decoding
that satisfy $\lim_{n\rarrow\infty}P_{err}=0$
for the CB net of
Eq.(\ref{eq-ch-net}).
\end{claim}
\proof
A proof
of this
is given in Ref.\cite{Tuc-redoing-classics}.
We
will not repeat that proof here.
That
proof
is very similar to the
achievability proofs that will be
presented later in this paper.
\qed

\begin{claim}\label{claim-ch-opt}
Optimality:
$\forall R_\rvm$, if $\exists$
an encoding and a decoding
that satisfy $\lim_{n\rarrow\infty}P_{err}=0$
for the CB net of
Eq.(\ref{eq-ch-net}),
then $R_\rvm\in \calr(\calp_{gen})$.
\end{claim}
\proof

A proof of this
is given in Ref.\cite{Tuc-redoing-classics}.
We
will not repeat that proof here.
That proof
is of the type most
commonly found in introductory
SIT textbooks.
Let us present here an
alternative
proof which is
more akin to
the optimality
proofs
which will be presented later in this paper.

\beqa
nR_\rvm &=& \ln N_\rvm = H(\rvm)
= H(\rvm:\rvy^n) + H(\rvm|\rvy^n)
\label{eq-ch-opt-a}
\\
&\leq &
H(\rvm:\rvy^n) + n\delta
\label{eq-ch-opt-b}
\\
&=&
\sum_{j=1}^n
H(\rvy_j:\rvy_{<j},\rvm)
+n\delta
\label{eq-ch-opt-c}
\\
&=&
nH(\rvy_\rvj:\rvy_{<\rvj},\rvm |\rvj)
+n\delta
\label{eq-ch-opt-d}
\\
&=&
n(H(\rvy:\rvu|\rvj)
+\delta)
\label{eq-ch-opt-e}
\\
&\leq &
n(H(\rvy:\rvu)
+\delta)
\label{eq-ch-opt-f}
\\
&\leq &
n(H(\rvy:\rvx)
+\delta)
\label{eq-ch-opt-g}
\;
\eeqa

\begin{itemize}
\item[(\ref{eq-ch-opt-b})]
This follows from Fano's inequality,
which says that if $\lim_{n\rarrow \infty}P_{err}=0$,
then $H(\rvm|\rvy^n)\leq n\delta$
with
$\lim_{n\rarrow \infty}\delta=0$.

\item[(\ref{eq-ch-opt-c})]
This follows from the chain rule for
MI.
\item[(\ref{eq-ch-opt-d})]
This follows from defining a
new, uniformly distributed
random variable $\rvj$
with states $j=1,2,\ldots,n$.

\item[(\ref{eq-ch-opt-e})]
This follows from a change
of notation $\rvy_\rvj\rarrow \rvy$
and $(\rvy_{<\rvj},\rvm,\rvj) \rarrow \rvu$.
\item[(\ref{eq-ch-opt-f})]
This follows
from Claim \ref{claim-cmi-leq-mi}.

\item[(\ref{eq-ch-opt-g})]
This follows from the classical data processing
inequalities.
\end{itemize}
\mbox{\;}
\qed

In the above optimality
proof, we
start with the CB net
of Eq.(\ref{eq-ch-net}).
Then we do some
``chain ruling"
reminiscent of
peeling away
all $n$ layers except
one.
We end up with
a different
CB net with
random variables $\rvy,\rvx,\rvu$.
It's instructive
to present
a chain of CB nets
connecting the
beginning and ending
CB nets of this process.

One starts with
the CB net ($\rv{\calc}$ implicit)
\begin{subequations}
\beq
\begin{array}{c}
\entrymodifiers={++[o][F-]}
\xymatrix{
*++[o][F=]{\what{\rvm}}&\rvy^n\ar[l]&\rvx^n\ar[l]&\rvm\ar[l]
}
\end{array}
\;.
\eeq
Tracing over (i.e., adding
over all states of)
the node $\rv{\what{m}}$
(highlighted with a double circle) gives

\beq
\begin{array}{c}
\entrymodifiers={++[o][F-]}
\xymatrix{
\rvy^n&\rvx^n\ar[l]&\rvm\ar[l]
}
\end{array}
\;.
\eeq
The previous CB net equals

\beq
\begin{array}{c}
\entrymodifiers={++[o][F-]}
\xymatrix{
\rvy_{<j}&
*+++[o][F=]{\rvx_{<j}}\ar[l]&*{}
\\
\rvy_j&\rvx_j\ar[l]&\rvm\ar[l]\ar[ul]\ar[dl]
\\
*++[o][F=]{\rvy_{>j}}&
*+++[o][F=]{\rvx_{>j}}\ar[l]&*{}
}
\end{array}
\;.
\eeq
Tracing over all
the nodes
highlighted with a double circle gives

\beq
\begin{array}{c}
\entrymodifiers={++[o][F-]}
\xymatrix{
*{}&\rvy_{<j}&*{}\\
\rvy_j&\rvx_j\ar[l]&\rvm\ar[l]\ar[ul]
}
\end{array}
\;.
\eeq
Merging the $\rvm$ and
$\rvy_j$ nodes
and promoting
the index $j$
to a
random variable, we get

\beq
\begin{array}{c}
\entrymodifiers={++[o][F-]}
\xymatrix{
\rvy_\rvj&\rvx_\rvj\ar[l]&*++++[o][F-]{
\begin{array}{c}
\rvu=
\\
{\scriptstyle\rvy_{<\rvj}\rvm}
\end{array}}\ar[l]
}
\end{array}
\;.
\eeq
\end{subequations}
In the last graph, we
leave implicit
a source node $\rvj$
with outgoing arrows pointing
into all nodes that
mention $\rvj$.

From Claims \ref{claim-ch-ach} and \ref{claim-ch-opt},
we see that $\calr(\calp_{gen})$
is the maximal achievable region
(MAR) of $R_\rvm$'s.
Some authors refer to the MAR as
the ``capacity region".
One defines the
channel capacity $C$ by

\beq
C= \max_{R_\rvm\in \calr(\calp_{gen})}R_\rvm
=\max_{P\in \calp_{gen}}H_P(\rvy:\rvx)
\;.
\eeq
Simply put,
for channel coding, the MAR is
just a closed interval $[0,C]$.
The upper limit point $C$ of this MAR is
called the channel capacity.

\section{Classical Wiretap Coding}

In this section, we
derive the channel
capacity for classical wiretap coding.
Everything
we prove in this
section has already
been proven by Csisz\'{a}r
and K\"{o}rner in Ref.\cite{CK}.
However,
some of our proofs (most
notably our achievability proof)
differ
significantly from
those presented in Ref.\cite{CK}.
We find that
our alternative proof methods are easier
to generalize
to quantum wiretap coding.

To help those readers who want
to compare our proofs to
those of Ref.\cite{CK},
we try in this paper
to use names of variables that
are identical
or at least very similar to those of
Ref.\cite{CK}.
As in Ref.\cite{CK},
the channel considered
in this paper
takes $x^n$ to
$(y^n,z^n)$.
Like Ref.\cite{CK},
we use the letter
$\rvs$ to denote the ``secret" signal,
one that is
visible only to
the observer that
measures the $y^n$
output. And
we
use the letter $\rvt$
to denote  the ``tapped"
or shared
signal, one that is
visible to both the $y^n$
and $z^n$ observers.

We consider all
wiretap coding protocols that can be
described by the following CB net

\beq
\begin{array}{c}
\entrymodifiers={++[o][F-]}
\xymatrix{
\what{\rvs}&\rvy^n\ar[dl]\ar[l]&*{}&\rvs\ar[dl]
\\
\what{\rvt}_\rvy&*{}
&\rvx^n\ar[ul]\ar[dl]&*{}
\\
\what{\rvt}_\rvz&\rvz^n\ar[l]\ar[uu]&*{}&\rvt\ar[ul]
\\
*{}&\rv{\calc}\ar[ruu]\ar[lu]\ar[luu]\ar[luuu]&*{}&*{}
}
\end{array}
\;
\label{eq-cla-net}
\eeq
with

\beq
P(s)=\frac{1}{N_\rvs}
\;,\;\;
P(t)=\frac{1}{N_\rvt}
\;,
\eeq

\beq
P(x^n|s,t,\calc)=\delta(x^n, x^n(s,t))
\;,
\eeq

\beq
P(y^n,z^n|x^n)=\prod_j P(y_j,z_j|x_j)
\;,
\eeq

\beq
P(\what{s}|y^n,\calc),
P(\what{t}_\rvy|y^n,\calc),
P(\what{t}_\rvz|z^n,\calc)=\mbox{ to be specified}
\;,
\eeq
and

\beq
P(\calc)=\mbox{to be specified}
\;.
\eeq
Assume that we are given
a channel $\{P(y,z|x)\}_{\forall y,z}
\in pd(S_{\rvy,\rvz})$
for all $x\in S_\rvx$.
The encoding $P(\calc)$
and
decoding $P(\what{s}|y^n,\calc)$,
$P(\what{t}_\rvy|y^n,\calc)$,
$P(\what{t}_\rvz|z^n,\calc)$
probability distributions are
yet to be specified.

We will consider 3 different
coding rates:

\beq
R_\rve=\frac{H(\rvs^k|\rvz^n)}{nk}
\;,\;\;
R_\rvs=\frac{\ln N_\rvs}{n}
\;,\;\;
R_\rvt=\frac{\ln N_\rvt}{n}
\;.
\eeq
By definition, all 3 R's
are non-negative.
To define $R_\rve$,
called the equivocation rate,
we've replaced
$\rvs$ by
a block of $k$
letters $\rvs^k$.
 CMI$\geq$ 0 implies that
$\frac{1}{nk}H(\rvs^k|\rvz^n)\leq
\sum_{j=1}^{k}\frac{1}{nk}H(\rvs_{j}|\rvz^n)$.
If we assume $H(\rvs_{j}|\rvz^n)$
is the same for all $j=1,2, \ldots k$, then
$\frac{1}{nk}H(\rvs^k|\rvz^n)\leq
\frac{1}{n}H(\rvs_1|\rvz^n)$.
Furthermore, MI$\geq$ 0
implies that
$\frac{1}{n}H(\rvs_1|\rvz^n)\leq
\frac{1}{n}H(\rvs_1)=R_\rvs$.
We can now set $\rvs_1$ to $\rvs$.
We have established that

\beq
0\leq R_\rve\leq
\frac{1}{n}H(\rvs|\rvz^n)\leq R_\rvs
\;.
\eeq

Note that
$R_\rve=R_\rvs$ implies
$H(\rvs)=H(\rvs|\rvz^n)$
or, equivalently, $H(\rvs:\rvz^n)=0$.
Thus the case $R_\rve=R_\rvs$
can be described as perfect secrecy.
The case $R_\rvt=0$
can be described as
zero tapping rate.

Let

\beq
\vec{R}=(R_\rve, R_\rvs,R_\rvt)
\;.
\eeq

We will
always assume in this paper that $\rvs$
and $\rvt$ are independent. (Ref.\cite{CK}
sometimes assumes they aren't).
Ref.\cite{CK}
defines $R_\rvs = \frac{H(\rvs|\rvt)}{n}$.
When
$\rvs$
and $\rvt$ are independent,
$H(\rvs|\rvt) = H(\rvs)=\ln N_\rvs$.

For a given, fixed channel
$P_{\rvy,\rvz|\rvx}$,
define

\beq
\calp_{gen} =
\left\{
P_{\rvy,\rvz,\rvv,\rvu}
\in
pd(S_{\rvy,\rvz,\rvv,\rvu}):
\begin{array}{l}
P(y,z,v,u)
=\sum_x
P(y,z|x)P(x|v)P(v|u)P(u)
,
\\
P_{\rvy,\rvz|\rvx}\mbox{ fixed}
\end{array}
\right\}
\;.
\eeq
In terms of CB nets,
the elements of $\calp_{gen}$
must have the following
graph topology

\beq
P_{\rvy,\rvz,\rvv,\rvu}=
\sum_x
\sandb{
\entrymodifiers={++[o][F-]}
\xymatrix{
\rvy&*{}&\rvv\ar[dl]\\
*{}&*++++[o][F-]{\rvx=x}\ar[ul]\ar[dl]&*{}\\
\rvz\ar[uu]&*{}&\rvu\ar[uu]
}
}
\;.
\label{eq-cla-calp-net}
\eeq
We will say that
$P(y,z|v,u)$ factors in $y$ and $z$ if
$P(y,z|v,u)=P(y|v,u)P(z|v,u)$
for all $y,z,v,u$. (i.e.,
conditional independence).
It is convenient to define the following
subset of $\calp_{gen}$

\beq
\calp_{fac}=
\{P\in \calp_{gen}: \mbox{ $P_{\rvy,\rvz|\rvv,\rvu}$ factors
in $y$ and $z$}
\}
\;.
\eeq
The elements of $\calp_{fac}$
have the same graph as Eq.(\ref{eq-cla-calp-net})
except without the arrow connecting $\rvz$ and $\rvy$.

For any
$P\in pd(S_{\rvy,\rvz,\rvv,\rvu})$,
we can define
a convex hull of
$\vec{R}$'s by

\beq
\calr(P)=
\left\{\vec{R}=(R_\rve, R_\rvs,R_\rvt)\in\RR^3:
\begin{array}{l}
0\leq R_\rve \leq R_\rvs, 0\leq R_\rvs, 0\leq R_\rvt,
\\
R_\rve\leq
H(\rvy:\rvv|\rvu)
-H(\rvz:\rvv|\rvu),
\\
R_\rvs + R_\rvt
\leq H(\rvy:\rvv|\rvu) +\ell,
\\
R_\rvt\leq \ell,
\\
\ell =\min\{H(\rvy:\rvu),H(\rvz:\rvu)\}
\\
\mbox{all $H$ evaluated at $P$}
\end{array}
\right\}
\;.
\eeq
It's also useful to consider, for any
$\calp\subset pd(S_{\rvy,\rvz,\rvv,\rvu})$, the
set

\beq
\calr(\calp)=
\bigcup_{P\in \calp} \calr(P)
\;.
\eeq
\begin{figure}[h]
    \begin{center}
    \epsfig{file=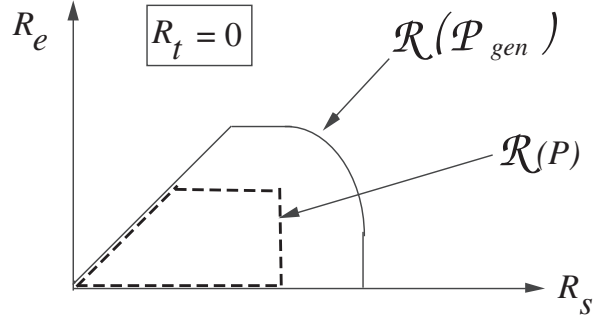, height=1.75in}
    \caption{$R_\rvt=0$
    crossections of the sets
    $\calr(\calp_{gen})$
    and $\calr(P)$
    for some particular point $P\in \calp_{gen}$.
    $\calr(P)$ is a convex hull,
    which is a set bounded
    by straight lines (or
    by planes in higher dimensions).
    $\calr(\calp_{gen})$ doesn't necessarily
    have completely straight sides but it must
    still be convex.
    }
    \label{fig-cla-mar}
    \end{center}
\end{figure}

$\calr(\calp)$
for arbitrary $\calp$
is not always a convex set.
For instance, it isn't if
$\calp$ consisting of just
two $P$'s.
But how about $\calr(\calp)$ when
$\calp=\calp_{gen}$?

\begin{claim}\label{claim-cla-convexity}
$\calr(\calp_{gen})$
is a closed convex set.
\end{claim}
\proof
$\calr(\calp_{gen})$
is a union of closed sets so it is closed.
Next we prove convexity.

Suppose $\rvb$
is a random variable with $S_\rvb=\{0,1\}$.
Let $E_b = \sum_b P(b)$.

In this proof, we will
use the shorthand

\beq
\rv{\xi}(b)=(\rvy(b),\rvz(b),\rvv(b),\rvu(b))
\;.
\eeq
for $b=0,1$.
Furthermore,
we will use $\rv{\xi}_j(b)$ with $j\in \{1,2,3,4\}$
to denote the four components of this vector.
$\rv{\xi}(0)$ and
$\rv{\xi}(1)$
represent two different CB nets. Note
that in general $S_{\rv{\xi}_j(0)}\neq
S_{\rv{\xi}_j(1)}$ for all $j$.

We want to prove that if
$\vec{R}(b)\in \calr(\calp_{gen})$
for $b=0,1$, then
$E_b\vec{R}(b)\in \calr(\calp_{gen})$.
This means that if we
are given 2 CB nets $\rv{\xi}(0)$
and $\rv{\xi}(1)$
such that
$P_{\rv{\xi}(b)}\in \calp_{gen}$
and
$\vec{R}(b)\in \calr(P_{\rv{\xi}(b)})$
for $b=0,1$,
the we can find an average CB net
$\rv{\xi}$ such that
$P_{\rv{\xi}}\in \calp_{gen}$
and
$E_b\vec{R}(b)\in \calr(P_{\rv{\xi}})$.
Define $\vec{R}= E_b \vec{R}(b)$.
We want to prove that
$\vec{R}\in \calr(P_{\rv{\xi}})$.

We define the average CB net
$\rv{\xi}$
to have the
topology prescribed by $\calp_{gen}$,
with nodes

\beq
\rv{\xi}_j=
[\rv{\xi}_j(0),\rv{\xi}_j(1),\rvb]
\;
\eeq
for all $j$.
If we assume that

\beq
P(\rv{\xi}=\xi|\rvb=b)
=
P(\rv{\xi}(b)=\xi(b))
\;,
\eeq
then

\beqa
P(\rv{\xi}=\xi)&=&
\sum_b P(\rv{\xi}=\xi|\rvb=b)P(b)
\\
&=&
\sum_b P(\rv{\xi}(b)=\xi(b))P(b)
\\
&=&
E_b
P(\rv{\xi}(b)=\xi(b))
\;.
\eeqa

One has
\beq
0\leq R_\rve(b)\leq R_\rvs(b)
\;
\eeq
for $b=0,1$.
Hence

\beq
0\leq E_b R_\rve(b)\leq E_b R_\rvs(b)
\;.
\eeq
Hence

\beq
0\leq R_\rve\leq R_\rvs
\;.
\eeq

Next note that
\beqa
\lefteqn{H_{P_{\rv{\xi}}}(\rvy:\rvv|\rvu)
-
H_{P_{\rv{\xi}}}(\rvy:\rvv|\rvu)
=}\nonumber
\\
&=&
H_{P_{\rv{\xi}}}(\rvy:\rvv|\rvu,\rvb)
-
H_{P_{\rv{\xi}}}(\rvy:\rvv|\rvu,\rvb)
\\
&=&
E_b \left\{H_{P_{\rv{\xi}}(b)}(\rvy(b):\rvv(b)|\rvu(b))
-
 H_{P_{\rv{\xi}}(b)}(\rvy(b):\rvv(b)|\rvu(b))\right\}
\\
&\geq&
E_b R_\rve(b)
\\
&=&
R_\rve
\;.
\eeqa

Next note that
\beqa
\lefteqn{H_{P_{\rv{\xi}}}(\rvy:\rvv|\rvu)
+
H_{P_{\rv{\xi}}}(\rvy:\rvu)
=}\nonumber
\\
&=&
H_{P_{\rv{\xi}}}(\rvy:\rvv|\rvu,\rvb)
+
H_{P_{\rv{\xi}}}(\rvy:\rvu,\rvb)
\\
&\geq&
H_{P_{\rv{\xi}}}(\rvy:\rvv|\rvu,\rvb)
+
H_{P_{\rv{\xi}}}(\rvy:\rvu|\rvb)
\\
&=&
E_b \left\{H_{P_{\rv{\xi}}(b)}(\rvy(b):\rvv(b)|\rvu(b))
+
H_{P_{\rv{\xi}}(b)}(\rvy(b):\rvu(b))\right\}
\\
&\geq&
E_b \{R_\rvs(b) + R_\rvt(b)\}
\\
&=&
R_\rvs + R_\rvt
\;.
\eeqa
Likewise,

\beq
H_{P_{\rv{\xi}}}(\rvy:\rvv|\rvu)
+
H_{P_{\rv{\xi}}}(\rvz:\rvu)
\geq
R_\rvs + R_\rvt
\;.
\eeq

Next note that
\beqa
H_{P_{\rv{\xi}}}(\rvy:\rvu)
&=&
H_{P_{\rv{\xi}}}(\rvy:\rvu,\rvb)
\\
&\geq&
H_{P_{\rv{\xi}}}(\rvy:\rvu|\rvb)
\\
&=&
E_b
H_{P_{\rv{\xi}}(b)}(\rvy(b):\rvu(b))
\\
&\geq&
E_b R_\rvt(b)
\\
&=&
R_\rvt
\;.
\eeqa
Likewise,

\beq
H_{P_{\rv{\xi}}}(\rvz:\rvu)
\geq
R_\rvt
\;.
\eeq
\qed

\begin{claim}
\beq
\calr(\calp_{gen})=
\calr(\calp_{fac})
\;.
\eeq
\end{claim}
\proof
$\calp_{fac}\subset \calp_{gen}$
so
$\calr(\calp_{fac})\subset \calr(\calp_{gen})$.
Next let's prove the reverse inclusion.
Suppose $P_{\rvy,\rvz,\rvv,\rvu}\in \calp_{gen}$.
$P_{\rvy,\rvz|\rvv,\rvu}$
only appears in the definition
of $\calr(P)$
through its marginals
$P_{\rvy|\rvv,\rvu}$
and
$P_{\rvz|\rvv,\rvu}$.
The set
$\calr(P)$
doesn't change if in its definition,
one
replaces
everywhere the
probability distribution
$P_{\rvy,\rvz|\rvv,\rvu}(y,z|v,u)$
by the product probability distribution
$P_{\rvy|\rvv,\rvu}(y|v,u)P_{\rvz|\rvv,\rvu}(z|v,u)$.
Both probability distributions
belong to $pd(S_{\rvy,\rvz})$
for all $v,u$.
Given $P\in \calp_{gen}$,
we have found
$\tilde{P}\in \calp_{fac}$
such that
$\calr(P)=\calr(\tilde{P})$.
\qed

\subsection{Optimality}
\label{sec-cla-opt}

\begin{claim}
Optimality:
$\forall \vec{R}$, if $\exists$
an encoding and a decoding
that satisfy $\lim_{n\rarrow\infty}P_{err}=0$
for the CB net of
Eq.(\ref{eq-cla-net}),
then
$\vec{R}\in \calr(\calp_{gen})$.
\end{claim}
\proof
The proof starts
from the result Claim \ref{claim-opt-identity}.
As pointed out in the section
that ends and culminates
with Claim \ref{claim-opt-identity},
Claim \ref{claim-opt-identity}
is basically a souped-up
version of the chain rule.
To prove the current claim,
we will need to add 3 new
ingredients that are
not consequences of
only the chain rule.
First, we will use Fano's inequality
(see Ref.\cite{CovTh}).
Second, we will use
the fact
that
classical clone random variables
can be merged. By this we mean that

\beq
H(\rva:\rvb, \rvc|\rvc)=H(\rva:\rvb|\rvc)
\;
\label{eq-cla-cloning}
\eeq
for any random variables $\rva,\rvb,\rvc$.
Third, we will use Claim
\ref{claim-cmi-leq-mi}.

Note that, by assumption,
$\lim_{n\rarrow \infty}P_{err}=0$.
Hence, by Fano's inequality,
the $\delta_y$, $\delta_{ty}$
and $\delta_{tz}$
used in Claim \ref{claim-opt-identity}
go to zero as $n\rarrow \infty$.
Furthermore, classical conditional
entropies are non-negative and CMI$\geq$0
so

\beq
0\leq \frac{1}{n}H(\rvt|\rvz^n,\rvs)\leq \frac{1}{n}H(\rvt|\rvz^n)
\;.
\eeq
But Fano's inequality implies that
$\lim_{n\rarrow \infty}\frac{1}{n}H(\rvt|\rvz^n)=0$
so also

\beq
\lim_{n\rarrow \infty}\frac{1}{n}H(\rvt|\rvz^n,\rvs)=0
\;.
\eeq

Let $\rvj$ be
a random variable
that is uniformly distributed
and has states $j=1,2,\cdots,n$.
Let $E_j=\frac{1}{n}\sum_j$.
Define

\beq
\begin{array}{l}
\rvu = (\rvy_{<\rvj},\rvz_{>\rvj},\rvt,\rvj),
\\
\rvv = (\rvu,\rvs),
\\
\rvx=\rvx_\rvj\;,\;\;
\rvy=\rvy_\rvj\;,\;\;
\rvz=\rvz_\rvj
\end{array}
\;.
\label{eq-cla-uv-notation}
\eeq

After dropping $\frac{1}{n}H(\rvt|\rvz^n,\rvs)$
and the deltas
$\delta_y,\delta_{ty},\delta_{tz}$,
the 4 identities
(souped-up chain rules)
 of Claim \ref{claim-opt-identity}
yield the following
4 inequalities.

\begin{itemize}
\item
Eq.(\ref{eq-opt-identity-re})
yields

\beqa
R_\rve
&\leq&\frac{1}{n}H(\rvs|\rvz^n)
\\
&=&
E_j
H(\rvy_j:\rvs|\rv{\alpha}_j,\rvt)
-
E_j
H(\rvz_j:\rvs|\rv{\alpha}_j,\rvt)
\\
&=&
H(\rvy_\rvj:\rvs|\rv{\alpha}_\rvj,\rvt,\rvj)
-
H(\rvz_\rvj:\rvs|\rv{\alpha}_\rvj,\rvt,\rvj)
\\
&=&
H(\rvy:\rvv|\rvu)
-H(\rvz:\rvv|\rvu)
\;.
\eeqa
\item
Eq.(\ref{eq-opt-identity-rs})
yields

\beqa
R_\rvs &=&
E_j
H(\rvy_j:\rvs|\rv{\alpha}_j,\rvt)
+\frac{\Sigma_{<\rvt}}{n}-
\frac{\Sigma_{<\rvs,\rvt}}{n}
\\
&=&
H(\rvy_\rvj:\rvs|\rv{\alpha}_\rvj,\rvt,\rvj)
+\frac{\Sigma_{<\rvt}}{n}-
\frac{\Sigma_{<\rvs,\rvt}}{n}
\\
&=&
H(\rvy:\rvv|\rvu)
+ \frac{\Sigma_{<\rvt}}{n}
- \frac{\Sigma_{<\rvs,\rvt}}{n}
\\
&\leq&
H(\rvy:\rvv|\rvu)
+ \frac{\Sigma_{<\rvt}}{n}
\;.
\eeqa
\item
Eq.(\ref{eq-opt-identity-rty})
yields

\beqa
R_\rvt
&=&
E_j
H(\rvy_j:\rv{\alpha}_j,\rvt)
-
E_j H(\rvy_j:\rvy_{<j})
-\frac{\Sigma_{<\rvt}}{n}
\\
&=&
H(\rvy_\rvj:\rv{\alpha}_\rvj,\rvt|\rvj)
-
H(\rvy_\rvj:\rvy_{<\rvj}|\rvj)
-\frac{\Sigma_{<\rvt}}{n}
\\
&=&
H(\rvy:\rvu|\rvj)
-H(\rvy_\rvj:\rvy_{<\rvj}|\rvj)
-\frac{\Sigma_{<\rvt}}{n}
\\
&\leq&
H(\rvy:\rvu|\rvj)
-\frac{\Sigma_{<\rvt}}{n}
\\
&\leq&
H(\rvy:\rvu)
-\frac{\Sigma_{<\rvt}}{n}
\;.
\label{eq-cla-rt-leq-y}
\eeqa
Eq.(\ref{eq-cla-rt-leq-y})
follows from Claim \ref{claim-cmi-leq-mi}.

\item
Eq.(\ref{eq-opt-identity-rtz})
yields

\beqa
R_\rvt
&=&
E_j
H(\rvz_j:\rv{\alpha}_j,\rvt)
-
E_j H(\rvz_j:\rvz_{>j})
-\frac{\Sigma_{<\rvt}}{n}
\\
&=&
H(\rvz_\rvj:\rv{\alpha}_\rvj,\rvt|\rvj)
-
H(\rvz_\rvj:\rvz_{>\rvj}|\rvj)
-\frac{\Sigma_{<\rvt}}{n}
\\
&=&
H(\rvz:\rvu|\rvj)
-H(\rvz_\rvj:\rvz_{>\rvj}|\rvj)
-\frac{\Sigma_{<\rvt}}{n}
\\
&\leq&
H(\rvz:\rvu|\rvj)
-\frac{\Sigma_{<\rvt}}{n}
\\
&\leq&
H(\rvz:\rvu)
-\frac{\Sigma_{<\rvt}}{n}
\;.
\label{eq-cla-rt-leq-z}
\eeqa
\end{itemize}

Eqs.(\ref{eq-cla-rt-leq-y})
and (\ref{eq-cla-rt-leq-z})
can be combined by writing

\beq
R_\rvt\leq \ell
-\frac{\Sigma_{<\rvt}}{n}
\;.
\eeq

From all we've said so far,
it's clear that $\vec{R}\in \calr(\calp_{gen})$.
\qed

In the above optimality
proof, we
start with the CB net
of Eq.(\ref{eq-cla-net}).
Then we do some
``chain ruling"
reminiscent of
peeling away
all $n$ layers except
one.
We end up with
a different
CB net with
random variables $\rvy,\rvz,\rvx,\rvv,\rvu$.
It's instructive
to present
a chain of CB nets
connecting the
beginning and ending
CB nets of this process.

One starts with
the CB net ($\rv{\calc}$ implicit)
\begin{subequations}
\beq
\begin{array}{c}
\entrymodifiers={++[o][F-]}
\xymatrix{
*+++[o][F=]{\what{\rvs}}&\rvy^n\ar[dl]\ar[l]&*{}&\rvs\ar[dl]
\\
*+++[o][F=]{\what{\rvt}_\rvy}&*{}
&\rvx^n\ar[ul]\ar[dl]&*{}
\\
*+++[o][F=]{\what{\rvt}_\rvz}&\rvz^n\ar[l]\ar[uu]&*{}&\rvt\ar[ul]
}
\end{array}
\;.
\eeq
Tracing over all
the nodes
highlighted with a double circle gives

\beq
\begin{array}{c}
\entrymodifiers={++[o][F-]}
\xymatrix{
\rvy^n&*{}&\rvs\ar[dl]
\\
*{}&\rvx^n\ar[ul]\ar[dl]&*{}
\\
\rvz^n\ar[uu]&*{}&\rvt\ar[ul]
}
\end{array}
\;.
\eeq
The previous CB net equals

\beq
\begin{array}{c}
\entrymodifiers={++[o][F-]}
\xymatrix@R=7pt{
\rvy_{<j}&*{}&*{}
\\
*{}&*+++[o][F=]{\rvx_{<j}}\ar[ul]\ar[dl]&\rvs\ar[l]\ar[lddd]\ar[ldddddd]
\\
*+++[o][F=]{\rvz_{<j}}\ar[uu]&*{}&*{}
\\
\rvy_j&*{}&*{}
\\
*{}&\rvx_j\ar[ul]\ar[dl]&*{}
\\
\rvz_j\ar[uu]&*{}&*{}
\\
*+++[o][F=]{\rvy_{>j}}&*{}&*{}
\\
*{}&*+++[o][F=]{\rvx_{>j}}\ar[ul]\ar[dl]&\rvt\ar[l]\ar[luuu]\ar[luuuuuu]
\\
\rvz_{>j}\ar[uu]&*{}&*{}
}
\end{array}
\;.
\eeq
Tracing over all
the nodes
highlighted with a double circle gives

\beq
\begin{array}{c}
\entrymodifiers={++[o][F-]}
\xymatrix@R=7pt{
\rvy_{<j}&*{}&\rvs\ar[ll]\ar[ldd]\ar@/^2pc/[ddddll]
\\
\rvy_j&*{}&*{}
\\
*{}&\rvx_j\ar[ul]\ar[dl]&*{}
\\
\rvz_j\ar[uu]&*{}&*{}
\\
\rvz_{>j}&*{}&\rvt\ar[ll]\ar[luu]\ar@/_2pc/[uuuull]
}
\end{array}
\;.
\eeq
Merging the $\rvy_{<j}$ and
$\rvz_{>j}$ nodes gives

\beq
\begin{array}{c}
\entrymodifiers={++[o][F-]}
\xymatrix{
\rvy_j&*{}&\rvs\ar[d]\ar[dl]
\\
*{}&\rvx_j\ar[ul]\ar[dl]&*++++[o][F-]{\scriptstyle\rvy_{<j}\rvz_{>j}}
\\
\rvz_j\ar[uu]&*{}&\rvt\ar[u]\ar[ul]
}
\end{array}
\;.
\eeq
The previous CB net
can be ``accommodated" or modeled
by the following CB net.

\beq
\begin{array}{c}
\entrymodifiers={++[o][F-]}
\xymatrix@R=7pt{
\rvy_\rvj&*{}&*+++[o][F-]{
\begin{array}{c}
\rvv=
\\ \scriptstyle\rvy_{<\rvj}\rvz_{>\rvj}\rvt\rvs
\end{array}
}\ar[dl]
\\
*{}&\rvx_\rvj\ar[ul]\ar[dl]&*{}
\\
\rvz_\rvj\ar[uu]&*{}&*++++[o][F-]{
\begin{array}{c}
\rvu=
\\
\scriptstyle\rvy_{<\rvj}\rvz_{>\rvj}\rvt
\end{array}}\ar[uu]
}
\end{array}
\;.
\eeq
\end{subequations}
In the last graph, we
leave implicit
a source node $\rvj$
with outgoing arrows pointing
into all nodes that
mention $\rvj$.

\subsection{Achievability}
\label{sec-cla-ach}

This section
will give a proof
of achievability for classical wiretap coding.
{\it The proof is very
similar to the one given
in Ref.\cite{Tuc-redoing-classics}
of achievability for classical channel
coding.}

\begin{claim}
Achievability:
$\forall \vec{R}$, if
$\vec{R}\in \calr(\calp_{gen})$, then
$\exists$
an encoding and a decoding
that satisfy $\lim_{n\rarrow\infty}P_{err}=0$
for the CB net of
Eq.(\ref{eq-cla-net}).
\end{claim}
\proof

Suppose  $P\in \calp_{gen}$
and $\vec{R}\in \calr(P)$.

We consider all
wiretap coding protocols that can be
described by the following CB net

\beq
\begin{array}{c}
\entrymodifiers={++[o][F-]}
\xymatrix{
\what{\rvs}&\rvy^n\ar[dl]\ar[l]&*{}&\rvv^n\ar[dl]&\rvs\ar[l]
\\
\what{\rvt}_\rvy&*{}
&\rvx^n\ar[ul]\ar[dl]&*{}&*{}
\\
\what{\rvt}_\rvz&\rvz^n\ar[l]\ar[uu]&*{}&\rvu^n\ar[uu]&\rvt\ar[l]
\\
*{}&\rv{\calc}\ar[ruu]\ar[rru]\ar[rruuu]\ar[lu]\ar[luu]\ar[luuu]&*{}&*{}&*{}
}
\end{array}
\;.
\label{eq-cla-extra-net}
\eeq

We will use a codebook
$\calc=(\calc_\rvx,\calc_\rvv,\calc_\rvu)$
composed
of
 3 sub-codebooks. Codebook
$\calc_\rvx$ is
as an $N_\rvs N_\rvt\times n $ matrix
given by
$\calc_\rvx = \{x^n(s,t)\}_{\forall s,t}= x^n(\cdot)$
where $x^n(s,t)\in S^n_\rvx$ for all $(s,t)
\in S_\rvs\times S_\rvt$.
Similarly,
$\calc_\rvv = \{v^n(s,t)\}_{\forall s,t}= v^n(\cdot)$
where $v^n(s,t)\in S^n_\rvv$ for all $(s,t)$
and $\calc_\rvu = \{u^n(t)\}_{\forall t}= u^n(\cdot)$
where $u^n(t)\in S^n_\rvu$ for all $t$.

We will use
the shorthand notations
$\what{M}= (\what{s},\what{t}_\rvy,\what{t}_\rvz)$
and
$m=(s,t)$. Let $f$ be the function
that maps
$f(m)=f(s,t) = (s,t,t)$
for all $s,t$.

We assign
to the CB net of Eq.(\ref{eq-cla-extra-net})
the following node transition matrices:

\beq
P(s)=\frac{1}{N_\rvs}
\;,\;\;
P(t)=\frac{1}{N_\rvt}
\;,
\eeq

\beq
P(u^n|t,\calc)=
\delta(u^n,u^n(t))
\;,
\eeq

\beq
P(v^n|u^n,s,\calc)
=
\delta(v^n,v^n(s,t))
\;,
\eeq
where $v^n(s,t)= v^n(s,u^n(t))$,

\beq
P(x^n|v^n,\calc)=\delta(x^n, x^n(s,t))
\;,
\eeq
where $x^n(s,t)=x^n(v^n(s,t))$,

\beq
P(y^n,z^n|x^n)=\prod_j P(y_j,z_j|x_j)
\;,
\eeq

\beq
P(\calc)=
\prod_{s,t}
P(x^n(s,t),v^n(s,t),u^n(t))
=
\prod_{s,t,j}
P_{\rvx,\rvv,\rvu}(x_j(s,t),v_j(s,t),u_j(t))
\;,
\eeq
where
$
P_{\rvx,\rvv,\rvu}(x,v,u)=
P_{\rvx|\rvv}(x|v)
P_{\rvv|\rvu}(v|u)
P_{\rvu}(u)
$,
and

\beq
P(\what{M}|
y^n,z^n,\calc)
=
\prod_{\mu=1}^5
\prod_{m_\mu\neq \what{m}_\mu}
\theta(R_\mu< \Gamma_\mu)
\;,
\eeq
where the quantities $\what{m}_\mu$,
$R_\mu$
and $\Gamma_\mu$
are defined in Appendix
\ref{ap-a}.
Assume that we are given
a channel $\{P(y,z|x)\}_{\forall y,z}
\in pd(S_{\rvy,\rvz})$
for all $x$.

The probability of success is defined by

\beq
P_{suc} =
P(\what{\rv{M}}= f(\rvm))
\;.
\eeq
One has

\beqa
P_{suc}
&=&
\sum_{\what{M},m}
\theta(\what{M}= f(m))
P(\what{M},m)
\\
&=&
\frac{1}{N_\rvs N_\rvt}
\sum_{\what{s},\what{t}}
\sum_\calc
P(\calc)
\sum_{y^n,z^n}
P(\what{M}=(\what{s},\what{t},\what{t})|y^n,z^n)
P(y^n,z^n|x^n(\what{s},\what{t}))
\\
&=&
E_\calc
\sum_{y^n,z^n}
P(y^n,z^n|x^n(\what{s},\what{t}))
\prod_{\mu=1}^5
\prod_{m_\mu\neq \what{m}_\mu}
\theta(R_\mu<\Gamma_\mu^{@\what{t}})
\;,
\label{eq-cla-psuc-pre-int_k}
\eeqa
where
$@\what{t}$ means evaluated
at $\what{t}_\rvy=\what{t}_\rvz=\what{t}$.

Let
\beq
\oint_k
=
\prod_{\mu=1}^5
\prod_{m_\mu\neq \what{m}_\mu}
\left\{
\int_{-\infty}^{+\infty}
\frac{dk_\mu(m_\mu)}{2\pi i}
\;\;
\frac{1}{(k_\mu(m_\mu)-i\eps)}
\right\}
\;,
\eeq
and

\beq
K_\mu = \sum_{m_\mu\neq \what{m}_\mu}
k_\mu(m_\mu)
\;.
\eeq
We will also use the following shorthand
notation

\beq
\what{x} = x(\what{s},\what{t})
\;,\;\;
\what{v} = v(\what{s},\what{t})
\;,\;\;
\what{u}= u(\what{t})
\;,
\eeq
and

\beq
\xi =
(y,z,x(\cdot), v(\cdot),u(\cdot))
\;.
\eeq
Expressing the $\theta$
functions in Eq.(\ref{eq-cla-psuc-pre-int_k})
 as
integrals,
we get

\beq
P_{suc}=
\oint_k
e^{-i\sum_\mu K_\mu R_\mu}
\sum_{y^n,z^n,x^n(\cdot), v^n(\cdot),u^n(\cdot)}
e^{
\sum_{\xi}
n\ptype{\;}
(\xi)
\ln Z(\xi)
}
\;,
\label{eq-cla-psuc-pre-p-type-int}
\eeq
where

\beqa
Z(\xi)&=&
P(y,z|\what{x})
\prod_{s,t}
\left\{
\begin{array}{c}P(x(s,t),v(s,t),u(t))\end{array}
\right\}
\prod_{\mu=1}^5
\prod_{m_\mu\neq \what{m}_\mu}
\left\{
(\gamma_\mu^{i\frac{k_\mu(m_\mu)}{n}})_{@\what{t}}
\right\}
\\
&=&
\left\{
\begin{array}{l}
P(y,z|\what{x})
\prod_{s,t}
\left\{
\begin{array}{c}P(x(s,t)|v(s,t),u(t))\end{array}
\right\}
\\
\prod_{s,t}
\left\{
\begin{array}{c}P(v(s,t),u(t))\end{array}
\right\}
\prod_{\mu=1}^5
\prod_{m_\mu\neq \what{m}_\mu}
\left\{
(\gamma_\mu^{i\frac{k_\mu(m_\mu)}{n}})_{@\what{t}}
\right\}
\end{array}\right.
\;,
\eeqa
where the quantities $\gamma_\mu$
are defined in Appendix \ref{ap-a}.

Next we express
the sum over $y^n,z^n,x^n(\cdot), v^n(\cdot),u^n(\cdot)$
in Eq.(\ref{eq-cla-psuc-pre-p-type-int})
as a p-type integral and
do the p-type integration
by the steepest descent method. We
get for the leading order term:

\beq
P_{suc}=
\oint_k
e^{-i\sum_\mu K_\mu R_\mu}
e^{n\ln Z}
\;,
\label{eq-cla-psuc-pre-delta}
\eeq
where

\beq
Z=
\sum_{\xi}
Z(\xi)
\;.
\label{eq-cla-xi-sum}
\eeq
Note that
in Eq.(\ref{eq-cla-xi-sum}),
the sum over the $x(\cdot)$
component of $\xi$
can be done easily because
none of the $\gamma_\mu$
depend on $x(\cdot)$
so we may
set

\beq
\sum_{x(\cdot)}
P(y,z|\what{x})
\prod_{s,t}
\left\{
P(x(s,t)|v(s,t),u(t))
\right\}
=
P(y,z|\what{v},\what{u})
\;.
\eeq
Hence,
using the shorthand notations

\beq
E_{y,z} =\sum_{y,z} P(y,z)
\;,
\eeq
and

\beq
E_{v(s,t),u(t)}=
\sum_{v(s,t), u(t)}P(v(s,t),u(t))
\;,\;\;
E_{v(\cdot),u(\cdot)}=
\prod_{s,t}E_{v(s,t),u(t)}
\;,
\eeq
$Z$ can be expressed as

\beq
Z=
E_{y,z}
E_{v(\cdot),u(\cdot)}
\left[
P(y,z:\what{v},\what{u})
\prod_{\mu=1}^5
\prod_{m_\mu\neq \what{m}_\mu}
\left\{
(\gamma_\mu^{i\frac{k_\mu(m_\mu)}{n}})_{@\what{t}}
\right\}
\right]
\;.
\label{eq-cla-z-exp}
\eeq

Define
\beq
Z_0
=
E_{y,z}
E_{\what{v},\what{u}}
\left[
P(y,z:\what{v},\what{u})
\prod_\mu\{\Phi_\mu^{i\frac{K_\mu}{n}}\}
\right]
\;,
\eeq
where the quantities $\Phi_\mu$
are defined in Appendix $\ref{ap-a}$.

Define
the integration operators

\beq
1_{h,K}=
\prod_\mu\left\{
\int_{-\infty}^{+\infty}dh_\mu\;
\int_{-\infty}^{+\infty}\frac{dK_\mu}{2\pi}\;
e^{-ih_\mu K_\mu }
\right\}
\;,
\eeq
and

\beq
1_{h>0,K}=
1_{h,K}\prod_\mu\theta(h_\mu>0)
\;.
\eeq

Note that 1 equals

\beqa
1 &=& \prod_\mu\left\{
\int_{-\infty}^{+\infty}dK_\mu\;
\delta(\sum_{m_\mu\neq \what{m}_\mu}
\left\{k_\mu(m_\mu)\right\}-K_\mu)
\right\}
\\
&=&
1_{h,K}
e^{i\sum_\mu h_\mu
\sum_{m_\mu\neq \what{m}_\mu}k_\mu(m_\mu)
}
\;.
\label{eq-cla-one-is}
\eeqa
Multiplying $P_{suc}$
by 1 certainly doesn't change it.
Thus
the right hand sides of
Eqs.(\ref{eq-cla-psuc-pre-delta})
and (\ref{eq-cla-one-is})
can be
multiplied to get

\beq
P_{suc}=
1_{h,K}
e^{-i\sum_\mu K_\mu R_\mu}
\oint_k
e^{i\sum_\mu h_\mu\sum_{m_\mu\neq \what{m}_\mu}k_\mu(m_\mu)}
e^{
n\ln Z
}
\;.
\label{eq-cla-pre-km-int}
\eeq

Next
we will assume that,
for all $m_\mu$,
when doing the contour
integration over $k_\mu(m_\mu)$
in Eq.(\ref{eq-cla-pre-km-int})
with $Z$ given by Eq.(\ref{eq-cla-z-exp}),
the
$e^{n\ln Z}$
can be evaluated at the value
$k_\mu(m_\mu)=i\eps\rarrow 0$ of the pole.
Symbolically, this means we will assume

\beqa
\oint_k
e^{i\sum_\mu
h_\mu\sum_{m_\mu\neq \what{m}_\mu}k_\mu(m_\mu)}
e^{n\ln Z}
&=&
e^{n\ln Z_0}
\oint_k
e^{i\sum_\mu
h_\mu\sum_{m_\mu\neq \what{m}_\mu}k_\mu(m_\mu)}
\\
&=&
e^{n\ln Z_0}
\prod_\mu\left\{
\theta(h_\mu >0)
\right\}
\;.
\label{eq-cla-magic-contour-int}
\eeqa
Applying Eq.(\ref{eq-cla-magic-contour-int})
to Eq.(\ref{eq-cla-pre-km-int}) gives

\beq
P_{suc}=
1_{h>0,K}
e^{-i\sum_\mu K_\mu R_\mu}
e^{
n\ln Z_0
}
\;.
\label{eq-cla-pre-exp-log}
\eeq
Expanding $\ln Z_0$
to first order in $K_\mu$
and doing the $h$ and $K$
integrals yields

\beqa
P_{suc} &\approx&
\prod_{\mu=1}^{5} \theta(R_\mu<H_\mu)
\\
&=& \theta(\vec{R}\in \calr(P))
\;,
\eeqa
where the quantities $H_\mu$
are defined in Appendix \ref{ap-a}.
\qed

\subsection{Capacity}
\label{sec-cla-cap}

From Sections \ref{sec-cla-opt} and
\ref{sec-cla-ach},
we see that $\calr(\calp_{gen})$
is the maximal achievable region
(MAR)
of $\vec{R}$'s.
The MAR is
a closed convex set.
We define the channel capacity
as the value of $R_\rvs$
at
one of the corners of the MAR.

Define the
line

\beq
\call =
\{\vec{R}\in \RR^3:
R_\rvt=0, R_\rve=R_\rvs
\}
\;.
\eeq
One defines the
channel capacity $C_\rvs$ by

\beq
C_\rvs=
\max_{R_\rvs\in \calr(\calp_{gen})\cap\call} R_\rvs
=
\max_{P\in \calp_{gen}}
\{
H_P(\rvy:\rvv|\rvu)
-H_P(\rvz:\rvv|\rvu)\}
\;.
\eeq

Some authors
put extra restrictions
on the wiretap channel
$P(y,z|x)$
and use Claim \ref{claim-markov-diff}
given below.
They do this in order
to reduce and
simplify the MAR
and to
simplify
the
formula for the
channel capacity.
It is also possible
to trim $\calp_{fac}$. That
is, to find $\calp\subset \calp_{fac}$
with $\calr(\calp)=\calr(\calp_{fac})$.
One can, for example,
place bounds on $N_\rvv$ and $N_\rvu$
by polynomial functions of $N_\rvx$.
This reduces the size of the space $\calp$
one must search over in order
to calculate the channel capacity.
See Ref.\cite{CK}
for more details.

\begin{claim}
\label{claim-markov-diff}
For a Markov chain
$\rva\larrow \rve\larrow \rvb$
(or any tri-node Markov-like chain
with $\rve$ in the middle),

\beq
H(\rva:\rve|\rvb)=
H(\rva:\rve)
-H(\rva:\rvb)
\;.
\label{eq-markov-diff}
\eeq
\end{claim}
\proof
Claim \ref{claim-swap-zero-one}
implies
\beq
H(\rva:\rve|\rvb)
=H(\rva:\rvb|\rve)
+ H(\rva:\rve)
-H(\rva:\rvb)
\;.
\eeq
But for the type of
graph that is
assumed as a premise,
one has $H(\rva:\rvb|\rve)=0$.
\qed

\section{Quantum Wiretap Coding}

In this section,
we consider
quantum wiretap coding.
We try to make
our treatment of
quantum wiretap coding
as parallel a possible
to our treatment
in the previous sections of
classical
wiretap coding.
This
parallel treatment
is facilitated
by the use
of CB nets
for the classical case and
QB nets for the quantum one.

We consider all
wiretap coding protocols that can be
described by the following QB net

\beq
\begin{array}{l}
\tr_{\rvr_\rvs}
\\
\tr_{\rvr_\rvt}
\end{array}
\sandb{
\entrymodifiers={++[o][F-]}
\xymatrix@R=7pt{
*{}&*{}&*{}&\rvr_\rvs&*{}
\\
\what{\rvs}&\cancel{\rvy^n}\ar[dl]\ar[l]&*{}
&*{}&\rvs\ar[dl]\ar[ul]
\\
\what{\rvt}_\rvy&*{}
&\scriptstyle\rvy^n,\rvz^n\ar[ul]_>>{\delta}\ar[dl]_>>{\delta}
&\rvx^n\ar[l]&*{}
\\
\what{\rvt}_\rvz&\cancel{\rvz^n}\ar[l]&*{}
&*{}&\rvt\ar[ul]\ar[dl]
\\
*{}&\rv{\calc}\ar[rruu]\ar[lu]\ar[luu]\ar[luuu]
&*{}&\rvr_\rvt&*{}
}
}
\sandb{\hc}
\;
\label{eq-qua-net}
\eeq
with

\beq
A(s)=\frac{1}{\sqrt{N_\rvs}}
\;,\;\;
A(t)=\frac{1}{\sqrt{N_\rvt}}
\;,
\eeq

\beq
A(x^n|s,t,\calc)=\delta(x^n, x^n(s,t))
\;,
\eeq

\beq
A(y^n,z^n|x^n)=\prod_j A(y_j,z_j|x_j)
\;,
\eeq

\beq
A(\what{s}|y^n,\calc),
A(\what{t}_\rvy|y^n,\calc),
A(\what{t}_\rvz|z^n,\calc)=
\mbox{ to be specified}
\;,
\eeq
and

\beq
A(\calc)=\mbox{to be specified}
\;.
\eeq
Assume that we are given
the reservoir amplitudes
$A(r_\rvs|s)$
and $A(r_\rvt|t)$.
Assume that we are also given
an \emph{isometry} $A(y,z|x)$ (called
the wiretap channel amplitude for this problem).
The encoding $A(\calc)$
and
decoding $A(\what{s}|y^n,\calc)$,
$A(\what{t}_\rvy|y^n,\calc)$,
$A(\what{t}_\rvz|z^n,\calc)$
probability amplitudes are
yet to be specified.

We will consider 3 different
coding rates:

\beq
R_\rve= \frac{S(\rvs^k|\rvz^n)}{nk}
\;,\;\;
R_\rvs=\frac{\ln N_\rvs}{n}
\;,\;\;
R_\rvt=\frac{\ln N_\rvt}{n}
\;.
\eeq
Clearly,
$R_\rvs$ and $R_\rvt$ must
be non-negative.
$R_\rve$ must be
non-negative too
even though it
is defined as a quantum
conditional entropy and those
can sometimes be negative.
But not this time,
at least not for very large $n$.
This is why. $-\frac{1}{n}S(\rvs^k|y^n)\leq
\frac{1}{n}S(\rvs^k|\rvz^n)$
by Claim \ref{claim-two-parents}.
Furthermore,
$\lim_{n\rarrow \infty}
-\frac{1}{n}S(\rvs^k|y^n)=0$
by Alicki-Fannes's inequality,
assuming an encoding and decoding
that satisfy $\lim_{n\rarrow \infty}P_{err}=0$.
Thus, $R_\rve\geq 0$.
To define $R_\rve$,
called the equivocation rate,
 we've replaced
$\rvs$ by
a block of $k$
letters $\rvs^k$.
 CMI$\geq$ 0 implies that
$\frac{1}{nk}S(\rvs^k|\rvz^n)\leq
\sum_{j=1}^{k}\frac{1}{nk}S(\rvs_{j}|\rvz^n)$.
If we assume $S(\rvs_{j}|\rvz^n)$
is the same for all $j=1,2, \ldots k$, then
$\frac{1}{nk}S(\rvs^k|\rvz^n)\leq
\frac{1}{n}S(\rvs_1|\rvz^n)$.
Furthermore, MI$\geq$ 0
implies that
$\frac{1}{n}S(\rvs_1|\rvz^n)\leq
\frac{1}{n}S(\rvs_1)=R_\rvs$.
We can now set $\rvs_1$ to $\rvs$.
We have established that

\beq
0\leq R_\rve\leq \frac{S(\rvs|\rvz^n)}{n}\leq R_\rvs
\;.
\eeq

Let

\beq
\vec{R}=(R_\rve, R_\rvs,R_\rvt)
\;.
\eeq

For a given, fixed channel
$A_{\rvy,\rvz|\rvx}$,
define

\beq
\calp_{gen}=
\left\{
\rho_{\rvy,\rvz,\rvv,\rvu}
\in
dm(\calh_{\rvy,\rvz,\rvv,\rvu})
:
\begin{array}{l}
\rho_{\rvy,\rvz,\rvv,\rvu}
=
\sum_r
\sandb{
\begin{array}{r}
\sum_{y,z,x,v,u}A(y,z|x)\ket{y}
\\
\ket{z}
\\
A(x|v,u)\;\;\;
\\
A(v|u)\ket{v}
\\
A(u)\ket{u}
\\
A(r|y,z,v,u)\;\;\;
\end{array}
}
\sandb{\hc}
\\
A_{\rvy,\rvz|\rvx}\mbox{ fixed}
\end{array}
\right\}
\;.
\eeq
In terms of QB nets,
the elements of $\calp_{gen}$
must have the following
graph topology

\beq
\rho_{\rvy,\rvz,\rvv,\rvu}
=
\tr_\rvr
\sandb{
\entrymodifiers={++[o][F-]}
\xymatrix{
\cancel{\rvy}&*{}&*{}&\rvv\ar[dl]
\\
*{}&
\scriptstyle\rvy,\rvz\ar[ul]_>>{\delta}\ar[dl]_>>{\delta}
&\cancel{x}\ar[l]&*{}
\\
\cancel{\rvz}&*{}&*{}&\rvu\ar[ul]\ar[uu]
}
\\
\rvr\larrow\rvy,\rvz,\rvv,\rvu
}
\sandb{\hc}
\;.
\label{eq-qua-calp-net}
\eeq
Note that, unlike
its classical
counterpart Eq.(\ref{eq-cla-calp-net}),
this graph has an
arrow from $\rvu$ to
$\rvx$.

Define

\beq
A(y,z|v,u)=
\sum_{x}
A(y,z|x)A(x|v,u)
\;.
\eeq
We will say that
$A(y,z|v,u)$ factors in $y$ and $z$ if
$A(y,z|v,u)=A(y|v,u)A(z|v,u)$
for all $y,z,v,u$. (i.e.,
conditional independence).
If also
$A(y|v,u)$ and $A(z|v,u)$
are both isometries, we
will say that
$A(y,z|v,u)$ iso-factors in $y$ and $z$.
It is convenient to define the following
subset of $\calp_{gen}$

\beq
\calp_{fac}=
\{P\in \calp_{gen}:
\mbox{ $A_{\rvy,\rvz|\rvv,\rvu}$ iso-factors
in $y$ and $z$}
\}
\;.
\eeq
The elements of $\calp_{fac}$
have the same graph as Eq.(\ref{eq-qua-calp-net})
because,
by convention,
the marginalizer nodes $\rvy$ and
$\rvz$
are drawn without an arrow
connecting them regardless of whether these
two nodes depend on each other or not.

For any
$\rho\in dm(\calh_{\rvy,\rvz,\rvv,\rvu})$,
we can define
a convex hull of
$\vec{R}$'s by

\beq
\calr(\rho)=
\left\{\vec{R}=(R_\rve, R_\rvs,R_\rvt)\in\RR^3:
\begin{array}{l}
0\leq R_\rve \leq R_\rvs, 0\leq R_\rvs, 0\leq R_\rvt,
\\
R_\rve\leq
S(\rvy:\rvv|\rvu)
-S(\rvz:\rvv|\rvu),
\\
R_\rvs + R_\rvt
\leq S(\rvy:\rvv|\rvu) +\ell,
\\
R_\rvt\leq \ell,
\\
\ell =\min\{S(\rvy:\rvu),S(\rvz:\rvu)\}
\\
\mbox{all $S$ evaluated at $\rho$}
\end{array}
\right\}
\;.
\eeq
It's also useful to consider, for any
$\calp\subset dm(\calh_{\rvy,\rvz,\rvv,\rvu})$,
the set

\beq
\calr(\calp)
=
\bigcup_{\rho\in\calp}
\calr(\rho)
\;.
\eeq

Fig.\ref{fig-cla-mar}
was given initially for classical
wiretap coding, but
it is
still valid for quantum wiretap coding.

\begin{claim}\label{claim-qua-convexity}
$\calr(\calp_{gen})$
is a closed convex set.
\end{claim}
\proof
The proof is almost identical
to that of
Claim \ref{claim-cla-convexity}.

The random variable $\rvb$
is still classical,
even in the quantum case.

Here are some small differences
between the 2 proofs.
Instead of
2 CB nets
with probability distributions
$P_{\rv{\xi}(b)}\in \calp_{gen}$,
we consider now
2 QB nets
with density matrices
$\rho_{\rv{\xi}(b)}\in \calp_{gen}$.
Instead of
an average CB net
with probability
distribution $P_{\rv{\xi}}=
E_b P_{\rv{\xi}(b)}$,
we consider
 a QB net
with density matrix
$\rho_{\rv{\xi}}=
E_b \rho_{\rv{\xi}(b)}$.
Instead of $H()$'s, we use $S()$'s.
\qed

\begin{claim}
\beq
\calr(\calp_{gen})=
\calr(\calp_{fac})
\;
\eeq
\end{claim}
\proof
We won't give a rigorous proof of this,
just a plausibility argument.

$\calp_{fac}\subset \calp_{gen}$
so
$\calr(\calp_{fac})\subset \calr(\calp_{gen})$.
Next let's prove the reverse inclusion.
Suppose $\rho_{\rvy,\rvz,\rvv,\rvu}\in \calp_{gen}$.
$\rho_{\rvy,\rvz,\rvv,\rvu}$
only appears in the definition
of $\calr(\rho)$
through its marginals
$\rho_{\rvy,\rvv,\rvu}$
and
$\rho_{\rvz,\rvv,\rvu}$.

For the rest of
this proof we will
use the shorthand notation $\xi=(v,u)$.

$\rho_{\rvz,\rvy,\rv{\xi}}$ has the form
\beq
\rho_{\rvz,\rvy,\rv{\xi}}=
\sum_r
\sandb{
\begin{array}{r}
\sum_{y,z,\xi}
A(y,z|\xi)\ket{y}
\\
A(r|y,z,\xi)\ket{z}
\\
A(\xi)\ket{\xi}
\end{array}
}
\sandb{\hc}
\;.
\eeq
We would like to
find a new density matrix
$\tilde{\rho}_{\rvz,\rvy,\rv{\xi}}$
in terms of the original
density matrix $\rho_{\rvz,\rvy,\rv{\xi}}$.
We would like
$\tilde{\rho}_{\rvz,\rvy,\rv{\xi}}$
to be of the form

\beq
\tilde{\rho}_{\rvz,\rvy,\rv{\xi}}
=
\sum_R
\sandb{
\begin{array}{r}
\sum_{y,z,\xi}
\tilde{A}(y|\xi)\tilde{A}(z|\xi)\ket{y}
\\
\tilde{A}(R|y,z,\xi)\ket{z}
\\
\tilde{A}(\xi)\ket{\xi}
\end{array}
}
\sandb{\hc}
\;,
\eeq
where
$\tilde{A}(y|\xi)$
and $\tilde{A}(z|\xi)$
are both isometries,
and such that
$\tilde{\rho}_{\rvz,\rvy,\rv{\xi}}$
and
$\rho_{\rvz,\rvy,\rv{\xi}}$
have the same marginal density matrices
$\rho_{\rvy,\rv{\xi}}$ and $\rho_{\rvz,\rv{\xi}}$.

We will
set $A(\xi)=\tilde{A}(\xi)$ for all $\xi$.
Stated more explicitly,
we want
$\tilde{\rho}_{\rvz,\rvy,\rv{\xi}}$
to satisfy the constraints:

\begin{subequations}
\beq
\sum_{R,y}
\sandb{
\begin{array}{r}
\sum_{z,\xi}
\tilde{A}(y|\xi)\tilde{A}(z|\xi)
\\
\tilde{A}(R|y,z,\xi)
\end{array}
}
\sandb{\hc
\\
z\rarrow z'
\\
\xi\rarrow \xi'}
=
\sum_{r,y}
\sandb{
\begin{array}{r}
\sum_{z,\xi}
A(y,z|\xi)
\\
A(r|y,z,\xi)
\end{array}
}
\sandb{\hc
\\
z\rarrow z'
\\
\xi\rarrow \xi'}
\yz
\;
\label{eq-cla-constraint-a}
\eeq
(which is equivalent to
$\tilde{\rho}_{\rvz\rv{\xi}}=
\rho_{\rvz\rv{\xi}}$
and
$
\tilde{\rho}_{\rvy,\rv{\xi}}
=
\rho_{\rvy,\rv{\xi}}$),

\beq
\sum_y
\sandb{\tilde{A}(y|\xi)}
\sandb{\hc
\\
\xi\rarrow\xi'
}
=
\delta_\xi^{\xi'}
\yz
\;,
\label{eq-cla-constraint-b}
\eeq
and

\beq
\sum_R
|\tilde{A}(R|y,z,\xi)|^2=1
\;.
\label{eq-cla-constraint-c}
\eeq
\end{subequations}

Let's count the number of
unknowns (i.e., real-valued
degrees of freedom).
\begin{itemize}
\item
$\tilde{A}(y|\xi)\in \CC$
so it contains $2N_\rvy N_{\rv{\xi}}$
unknowns.
\item
$\tilde{A}(z|\xi)\in \CC$
so it contains
$2N_\rvz N_{\rv{\xi}}$ unknowns.
\item
$\tilde{A}(R|y,z,\xi))\in \CC$
so it contains
$2N_{\rv{R}}N_\rvy N_\rvz N_{\rv{\xi}}$
unknowns.
\end{itemize}
That's a total of
$$2N_\rvy N_{\rv{\xi}} +
2N_\rvz N_{\rv{\xi}} +
2N_{\rv{R}}N_\rvy N_\rvz N_{\rv{\xi}}$$
unknowns.

Next, let's count the number of
independent equations
(i.e., independent real-valued constraints).
Recall that an $N\times N$ Hermitian
matrix contains
$N^2$ degrees of freedom.\footnote{Let the
Hermitian matrix
have entries
$H_{i,j}$. If
$i< j$, store the real part of $H_{i,j}$
at position $(i,j)$
and the imaginary part at position $(j,i)$.
If $i=j$, $H_{i,i}$ is real. Store it at
position $(i,i)$.}
\begin{itemize}
\item
Eqs.(\ref{eq-cla-constraint-a})
give $(N_\rvz N_{\rv{\xi}})^2 + (N_\rvy N_{\rv{\xi}})^2$
independent equations.

\item
Eqs.(\ref{eq-cla-constraint-b})
give $2N_{\rv{\xi}}^2$
independent equations.
\item
Eqs.(\ref{eq-cla-constraint-c})
give $N_\rvy N_\rvz N_{\rv{\xi}}$
independent equations.
\end{itemize}

That's a total
of
$$(N_\rvz N_{\rv{\xi}})^2 + (N_\rvy N_{\rv{\xi}})^2
+
2N_{\rv{\xi}}^2
+
N_\rvy N_\rvz N_{\rv{\xi}}
$$
independent equations.

Now note that
we can always make $N_{\rv{R}}$
large enough so that the number
of unknowns is greater or
equal to the number of
equations. The moral is that when
we replace $A(y,z|\xi)$
by a product of two isometries
$\tilde{A}(y|\xi)$ and $\tilde{A}(z|\xi)$,
we are losing degrees of freedom.
Those extra degrees of freedom
can be transferred to the reservoir
whenever
the only part of
$\rho_{\rvy,\rvz,\rvv,\rvu}$
that is visible
are
the marginal density matrices
$\rho_{\rvy,\rvv,\rvu}$
and $\rho_{\rvz,\rvv,\rvu}$.

Given $\rho\in \calp_{gen}$,
we have found
$\tilde{\rho}\in \calp_{fac}$
such that
$\calr(\rho)=\calr(\tilde{\rho})$.
\qed

For quantum wiretap
coding, we will
prove below an optimality theorem
for $\calr(\calp_{gen})$
and an achievability theorem
for $\calr(\calp_{fac})$.
Since $\calr(\calp_{gen})=
\calr(\calp_{fac})$, this will show
achievability and optimality
over the same set of $\vec{R}$'s.

\subsection{Measuring Success in Quantum Communication}

When considering communication
through any channel,
whether it be a classical
or a quantum one,
it behooves us to
define a measure
of the success of that
communication.
In this section, we
will review the
measures of success in communication
that we used
previously for classical channel coding and
classical wiretap coding. Then
we will explain how we will
measure success in communication
for quantum wiretap coding.

In classical channel coding,
an $N_\rvm\times N_\rvm$
transition matrix $\{P(\what{m}|m)\}_{\forall \what{m},m}$
sends $P_\rvm\in pd(S_\rvm)$
to $P_{\rv{\what{m}}}\in pd(S_\rvm)$
as follows
\beq
P_{\rv{\what{m}}}(\what{m})
=
\sum_{m}P(\what{m}|m)P_\rvm(m)
\;,
\eeq
where

\beq
P(\what{m}|m)
=\sum_{y^n}
P(\what{m}|y^n)P(y^n|x^n(m))
\;.
\eeq
$P_{err}$
was defined in Ref.\cite{Tuc-redoing-classics}
as the sum of
the off-diagonal entries of the
transition matrix
divided by $N_\rvm$.
Thus, $P_{err}=0$
if and only if
$P(\what{m}|m)=\delta_m^{\what{m}}$.
We found that this occurs for small enough
values of $R_\rvm$:

\beq
P(\what{m}|m)
=
\delta(\what{m},m)
\theta(R_\rvm\leq H(\rvy:\rvx))
\;.
\eeq
Note that our definition
of $P_{err}$,
as the sum of the off-diagonal entries
of $P(\what{m}|m)/N_\rvm$, is
just one possible
definition of a distance
between
$P(\what{m}|m)$ and $\delta_m^{\what{m}}$.
We could, for example,
have chosen
instead, as a measure of that distance,
the maximum
of the off diagonal entries
of $P(\what{m}|m)$.

In classical wiretap coding,
an $N_{\rv{\what{M}}}\times N_\rvm$
 transition matrix $\{P(\what{M}|m)\}_{\forall \what{M},m}$
sends $P_\rvm\in pd(S_\rvm)$
to $P_{\rv{\what{M}}}\in pd(S_{\rv{\what{M}}})$
as follows

\beq
P_{\rv{\what{M}}}(\what{M})
=
\sum_{m}P(\what{M}|m)P_\rvm(m)
\;,
\eeq
where

\beq
P(\what{M}|m)
=\sum_{y^n,z^n}
P(\what{M}|y^n,z^n)P(y^n,z^n|x^n(m))
\;.
\eeq
Let
the non-diagonal entries
of the transition
matrix be defined as
those entries
whose indices $\what{M},m$
satisfy $\what{M}\neq m$.
$P_{err}$
was defined
earlier in this paper as the sum of
the non-diagonal entries
of
the transition matrix
divided by $N_\rvm$.
Thus, $P_{err}=0$
iff $P(\what{M}|m)=\delta(\what{M},f(m))$.
We found that this occurs for small enough
values of $\vec{R}$:

\beq
P(\what{M}|m)
=
\delta(\what{M},f(m))
\theta(\vec{R}\in \calr(P))
\;.
\eeq
Note that our definition
of $P_{err}$
is
just one possible
definition of a distance
between
$P(\what{M}|m)$ and $\delta(\what{M},f(m))$.

In quantum wiretap coding,
a channel superoperator
$\calt_{\rv{\what{M}}|\rvm}$
sends
a density matrix
$\rho^{in}_\rvm\in dm(\calh_\rvm)$
into a density
matrix $\rho^{out}_{\rv{\what{M}}}
\in dm(\calh_{\rv{\what{M}}})$
as follows

\beq
\rho^{out}_{\rv{\what{M}}}
=
\calt_{\rv{\what{M}}|\rvm}
(\rho^{in}_\rvm)
\;.
\label{eq-rho-in-to-out}
\eeq
In general,
the density matrix $\rho^{in}_\rvm$
has
matrix elements of the form:

\beq
\av{m|\rho^{in}_\rvm|m'}
=
\sum_{r_\rvm}
\sandb{A(r_\rvm|m)A(m)}
\sandb{\hc \\m\rarrow m'}
\;
\eeq
for all $m,m'\in S_\rvm$.
One can also find the
matrix elements of the
channel
superoperator
$\calt_{\rv{\what{M}}|\rvm}$.
They have two indices
$\what{M},\what{M'}\in S_\rv{\what{M}}$
 and two indices $m,m'\in S_\rvm$.
From Eq.(\ref{eq-rho-in-to-out})
one gets

\beqa
\av{\what{M}|
\rho^{out}_{\rv{\what{M}}}|
\what{M}'}
&=&
\sum_{m,m'}
\bra{\what{M}}
\calt_{\rv{\what{M}}|\rvm}
\left(\ket{m}\av{m|\rho^{in}_\rvm|m'}\bra{m'}\right)
\ket{\what{M}'}
\\
&=&
\sum_{m,m'}\av{m|\rho^{in}_\rvm|m'}
\;\;\;
\bra{\what{M}}
\calt_{\rv{\what{M}}|\rvm}
\left(\ket{m}\bra{m'}\right)
\ket{\what{M}'}
\;
\eeqa
where the
matrix elements
$\bra{\what{M}}
\calt_{\rv{\what{M}}|\rvm}
\left(\ket{m}\bra{m'}\right)
\ket{\what{M}'}$
of the channel
superoperator
$\calt_{\rv{\what{M}}|\rvm}$
have the general form

\beq
\bra{\what{M}}
\calt_{\rv{\what{M}}|\rvm}
\left(\ket{m}\bra{m'}\right)
\ket{\what{M}'}
=
\sum_{y^n,z^n}
\sandb{A(\what{M},y^n,z^n|m)}
\sandb{\hc\\
\what{M}\rarrow \what{M}'\\
m\rarrow m'
}
\;.
\eeq
For quantum wiretap
 coding,
 rather than
give one of many
possible definitions
of $P_{err}$,
we shall prove below
 that:
the matrix elements of the channel
superoperator
equal
$\delta(\what{M}, f(m))
\delta(\what{M'}, f(m'))$
for small enough $\vec{R}$'s. That is,
we will prove that

\beq
\bra{\what{M}}
\calt_{\rv{\what{M}}|\rvm}
\left(\ket{m}\bra{m'}\right)
\ket{\what{M}'}
=
\delta(\what{M}, f(m))
\delta(\what{M'}, f(m'))
\theta(\vec{R}\in \calr(\rho))
\;.
\eeq
Thus, the channel superoperator
$\calt_{\rv{\what{M}}|\rvm}$
can transmit
density
matrices
$\rho_\rvm^{in}$
faithfully
for small enough
$\vec{R}$'s.

\subsection{Optimality}

\begin{claim}\label{claim-opt-qua}
Optimality:
$\forall \vec{R}$, if $\exists$
an encoding and a decoding
that satisfy $\lim_{n\rarrow\infty}P_{err}=0$
for the QB net of
Eq.(\ref{eq-qua-net}),
then
$\vec{R}\in \calr(\calp_{gen})$.
\end{claim}
\proof
The proof of optimality
for quantum wiretap coding
is very similar to the
proof of optimality for
classical wiretap coding.
The main difference
is that Eq.(\ref{eq-cla-cloning})
cannot be used in the quantum case.
In the quantum case,
random variables cannot
be cloned unless they
are classical.(See Ref.\cite{Tuc-inequalities}
for a discussion of this).
Because of this,
we will end up defining
$\rvv$ and $\rvu$
differently
for quantum and classical
wiretap coding.

The proof starts
from the result Claim \ref{claim-opt-identity}.
As pointed out in the section
that ends and culminates
with Claim \ref{claim-opt-identity},
Claim \ref{claim-opt-identity}
is basically a souped-up
version of the chain rule.
To prove the current claim,
we will need to add 2 new
ingredients that are
not consequences of
only the chain rule.
First, we will use Alicki Fannes's inequality
(see Ref.\cite{Wilde}), which
is a generalization
to the quantum realm
of Fano's inequality.
Second, we will  use Claim
\ref{claim-cmi-leq-mi}
with $H()\rarrow S()$.
This claim
is still valid in the quantum
realm as long as $\rvj$ is
a classical random variable.

Note that, by assumption,
$\lim_{n\rarrow \infty}P_{err}=0$.
Hence, by Alicki Fannes's inequality,
the $\delta_y$,$\delta_{ty}$
and $\delta_{tz}$
used in Claim \ref{claim-opt-identity}
go to zero as $n\rarrow \infty$.
Furthermore, Claim
\ref{claim-two-parents} and CMI$\geq$0
imply that

\beq
-\frac{1}{n}S(\rvt|y^n)\leq
\frac{1}{n}S(\rvt|\rvz^n,\rvs)\leq
\frac{1}{n}S(\rvt|\rvz^n)
\;.
\eeq
But Alicki Fannes's inequality implies that
both sides of the last equation,
$-\frac{1}{n}S(\rvt|y^n)$ and
$\frac{1}{n}S(\rvt|\rvz^n)$,
 tend
to zero as $n\rarrow \infty$.
Hence, also

\beq
\lim_{n\rarrow \infty}
\frac{1}{n}S(\rvt|\rvz^n,\rvs)=0
\;.
\eeq

Let $\rvj$ be
a {\it classical} random variable
that is uniformly distributed
and has states $j=1,2,\cdots,n$.
Let $E_j=\frac{1}{n}\sum_j$.
In the quantum case,
we define

\beq
\begin{array}{l}
\rvu = (\rvy_{<\rvj},\rvz_{>\rvj},\rvt,\rvj),
\\
\rvv = \rvs,
\\
\rvx=\rvx_\rvj\;,\;\;
\rvy=\rvy_\rvj\;,\;\;
\rvz=\rvz_\rvj
\end{array}
\;.
\label{eq-qua-uv-notation}
\eeq
(Compare this
with Eq.(\ref{eq-cla-uv-notation})
for the classical case).

After dropping $\frac{1}{n}S(\rvt|\rvz^n,\rvs)$
and the deltas
$\delta_y,\delta_{ty},\delta_{tz}$,
the 4 identities
(souped-up chain rules)
 of Claim \ref{claim-opt-identity}
yield the following
4 inequalities .
\begin{itemize}
\item
Eq.(\ref{eq-opt-identity-re})
yields

\beqa
R_\rve
&\leq&
\frac{1}{n}S(\rvs|\rvz^n)
\\
&=&
E_j
S(\rvy_j:\rvs|\rv{\alpha}_j,\rvt)
-
E_j
S(\rvz_j:\rvs|\rv{\alpha}_j,\rvt)
\\
&=&
S(\rvy_\rvj:\rvs|\rv{\alpha}_\rvj,\rvt,\rvj)
-
S(\rvz_\rvj:\rvs|\rv{\alpha}_\rvj,\rvt,\rvj)
\\
&=&
S(\rvy:\rvv|\rvu)
-S(\rvz:\rvv|\rvu)
\;.
\eeqa
\item
Eq.(\ref{eq-opt-identity-rs})
yields

\beqa
R_\rvs
&=&
E_j
S(\rvy_j:\rvs|\rv{\alpha}_j,\rvt)
+\frac{\Sigma_{<\rvt}}{n}-
\frac{\Sigma_{<\rvs,\rvt}}{n}
\\
&=&
S(\rvy_\rvj:\rvs|\rv{\alpha}_\rvj,\rvt,\rvj)
+\frac{\Sigma_{<\rvt}}{n}-
\frac{\Sigma_{<\rvs,\rvt}}{n}
\\
&=&
S(\rvy:\rvv|\rvu)
+ \frac{\Sigma_{<\rvt}}{n}
- \frac{\Sigma_{<\rvs,\rvt}}{n}
\\
&\leq&
S(\rvy:\rvv|\rvu)
+ \frac{\Sigma_{<\rvt}}{n}
\;.
\eeqa
\item
Eq.(\ref{eq-opt-identity-rty})
yields

\beqa
R_\rvt
&=&
E_j
S(\rvy_j:\rv{\alpha}_j,\rvt)
-
E_j S(\rvy_j:\rvy_{<j})
-\frac{\Sigma_{<\rvt}}{n}
\\
&=&
S(\rvy_\rvj:\rv{\alpha}_\rvj,\rvt|\rvj)
-
S(\rvy_\rvj:\rvy_{<\rvj}|\rvj)
-\frac{\Sigma_{<\rvt}}{n}
\\
&=&
S(\rvy:\rvu|\rvj)
-S(\rvy_\rvj:\rvy_{<\rvj}|\rvj)
-\frac{\Sigma_{<\rvt}}{n}
\\
&\leq&
S(\rvy:\rvu|\rvj)
-\frac{\Sigma_{<\rvt}}{n}
\\
&\leq&
S(\rvy:\rvu)
-\frac{\Sigma_{<\rvt}}{n}
\;.
\label{eq-qua-rt-leq-y}
\eeqa
Note that $\rvj$
can be cloned because it's classical.
Eq.(\ref{eq-qua-rt-leq-y})
follows from Claim \ref{claim-cmi-leq-mi}.

\item
Eq.(\ref{eq-opt-identity-rtz})
yields

\beqa
R_\rvt
&=&
E_j
S(\rvz_j:\rv{\alpha}_j,\rvt)
-
E_j S(\rvz_j:\rvz_{>j})
-\frac{\Sigma_{<\rvt}}{n}
\\
&=&
S(\rvz_\rvj:\rv{\alpha}_\rvj,\rvt|\rvj)
-
S(\rvz_\rvj:\rvz_{>\rvj}|\rvj)
-\frac{\Sigma_{<\rvt}}{n}
\\
&=&
S(\rvz:\rvu|\rvj)
-S(\rvz_\rvj:\rvz_{>\rvj}|\rvj)
-\frac{\Sigma_{<\rvt}}{n}
\\
&\leq&
S(\rvz:\rvu|\rvj)
-\frac{\Sigma_{<\rvt}}{n}
\\
&\leq&
S(\rvz:\rvu)
-\frac{\Sigma_{<\rvt}}{n}
\;.
\label{eq-qua-rt-leq-z}
\eeqa
\end{itemize}

Eqs.(\ref{eq-qua-rt-leq-y})
and (\ref{eq-qua-rt-leq-z})
can be combined by writing

\beq
R_\rvt\leq \ell
-\frac{\Sigma_{<\rvt}}{n}
\;.
\eeq

From all we've said so far,
it's clear that $\vec{R}\in \calr(\calp_{gen})$.
\qed

In the above optimality
proof, we
start with the QB net
of Eq.(\ref{eq-qua-net}).
Then we do some
``chain ruling"
reminiscent of
peeling
$n-1$ layers away
down to just
one.
We end up with
a different
QB net with
random variables $\rvy,\rvz,\rvx,\rvv,\rvu$.
It's instructive
to present
a chain of QB nets
connecting the
beginning and ending
QB nets of this process.

One starts with
the QB net ($\rv{\calc}$ implicit)
\begin{subequations}
\beq
\sandb{
\entrymodifiers={++[o][F-]}
\xymatrix{
*+++[o][F=]{\what{\rvs}}&\rvy^n\ar[dl]\ar[l]
&*{}&*{}&\rvs\ar[dl]
\\
*+++[o][F=]{\what{\rvt}_\rvy}&*{}&
\cancel{\scriptstyle \rvy^n,\rvz^n}\ar[ul]_>>{\delta}\ar[dl]_>>{\delta}
&\rvx^n\ar[l]
\\
*+++[o][F=]{\what{\rvt}_\rvz}&\rvz^n\ar[l]
&*{}&*{}&\rvt\ar[ul]
}}
\sandb{\hc}
\;.
\eeq
Tracing over all
the nodes
highlighted with a double circle
and lumping the
traced nodes into
a reservoir $\rvr$ gives

\beq
\sandb{
\entrymodifiers={++[o][F-]}
\xymatrix{
\rvy^n
&*{}&*{}&\rvs\ar[dl]
\\
*{}&
\cancel{\scriptstyle \rvy^n,\rvz^n}\ar[ul]_>>{\delta}\ar[dl]_>>{\delta}
&\rvx^n\ar[l]
\\
\rvz^n
&*{}&*{}&\rvt\ar[ul]
}
\\
\rvr\larrow \rvy^n, \rvz^n}
\sandb{\hc}
\;.
\eeq
The previous QB net equals

\beq
\sandb{
\entrymodifiers={++[o][F-]}
\xymatrix@R=7pt{
\rvy_{<j}&*{}&*{}&*{}
\\
*{}&\cancel{\scriptstyle\rvy_{<j}\rvz_{<j}}\ar[ul]_>>{\delta}\ar[dl]_>>{\delta}
&*+++[o][F=]{\rvx_{<j}}\ar[l]&\rvs\ar[l]\ar[lddd]\ar[ldddddd]
\\
*+++[o][F=]{\rvz_{<j}}&*{}&*{}&*{}
\\
\rvy_j&*{}&*{}
\\
*{}&\cancel{\scriptstyle\rvy_{j}\rvz_{j}}\ar[ul]_>>{\delta}\ar[dl]_>>{\delta}
&\rvx_j\ar[l]&*{}
\\
\rvz_j&*{}&*{}&*{}
\\
*+++[o][F=]{\rvy_{>j}}&*{}&*{}&*{}
\\
*{}&\cancel{\scriptstyle\rvy_{>j}\rvz_{>j}}\ar[ul]_>>{\delta}\ar[dl]_>>{\delta}
&*+++[o][F=]{\rvx_{>j}}\ar[l]&\rvt\ar[l]\ar[luuu]\ar[luuuuuu]
\\
\rvz_{>j}&*{}&*{}&*{}
}
\\
\rvr\larrow \rvy^n, \rvz^n}
\sandb{\hc}
\;.
\eeq
Tracing over all
the nodes
highlighted with a double circle gives

\beq
\sandb{
\entrymodifiers={++[o][F-]}
\xymatrix@R=7pt{
\rvy_{<j}&*{}&*{}&\rvs\ar[lll]\ar[ldd]
\ar@/^3.2pc/[ddddlll]
\\
\rvy_j&*{}&*{}&*{}
\\
*{}&\cancel{\scriptstyle\rvy_j,\rvz_j}
\ar[ul]_>>{\delta}\ar[dl]_>>{\delta}&\rvx_j\ar[l]&*{}
\\
\rvz_j&*{}&*{}
\\
\rvz_{>j}&*{}&*{}&\rvt\ar[lll]\ar[luu]
\ar@/_3.2pc/[uuuulll]
}
\\
\rvr\larrow \rvy_j, \rvz_j, \rvy_{<j},\rvz_{>j}}
\sandb{\hc}
\;.
\eeq
Merging the $\rvy_{<j}$ and
$\rvz_{>j}$ nodes gives

\beq
\sandb{
\entrymodifiers={++[o][F-]}
\xymatrix{
\rvy_j&*{}&*{}&\rvs\ar[d]\ar[dl]
\\
*{}&\cancel{\scriptstyle\rvy_j,\rvz_j}\ar[ul]_>>{\delta}\ar[dl]_>>{\delta}
&\rvx_j\ar[l]&*++++[o][F-]{\scriptstyle\rvy_{<j}\rvz_{>j}}
\\
\rvz_j&*{}&*{}&\rvt\ar[u]\ar[ul]
}
\\
\rvr\larrow \rvy_j, \rvz_j, \rvy_{<j},\rvz_{>j}}
\sandb{\hc}
\;.
\eeq
The previous QB net
can be ``accommodated" or modeled
by the following QB net.

\beq
\sandb{
\entrymodifiers={++[o][F-]}
\xymatrix@R=7pt{
\rvy_\rvj&*{}&*{}&*+++[o][F-]{
\begin{array}{c}
\rvv=
\\ \rvs
\end{array}
}\ar[dl]
\\
*{}&\cancel{\scriptstyle\rvy_\rvj,\rvz_\rvj}\ar[ul]_>>{\delta}\ar[dl]_>>{\delta}
&\rvx_\rvj\ar[l]&*{}
\\
\rvz_\rvj&*{}&*{}&*++++[o][F-]{
\begin{array}{c}
\rvu=
\\
\scriptstyle\rvy_{<\rvj}\rvz_{>\rvj}\rvt
\end{array}}\ar[uu]\ar[ul]
}
\\
\rvr\larrow \rvy_\rvj, \rvz_\rvj, \rvv,\rvu}
\sandb{\hc}
\;.
\label{eq-qua-net-opt}
\eeq
\end{subequations}
In the last graph, we
leave implicit
a (classical) source node $\rvj$
with outgoing arrows pointing
into all nodes that
mention $\rvj$.

Note that in Claim \ref{claim-opt-qua}
(the optimality proof for quantum wiretap
coding)
all the
quantum entropies $S()$
were evaluated
for the density matrix Eq.(\ref{eq-qua-net-opt}).

\subsection{Achievability}

\begin{claim}
Achievability:
$\forall \vec{R}$, if
$\vec{R}\in \calr(\calp_{gen})$, then
$\exists$
an encoding and a decoding
that satisfy $\lim_{n\rarrow\infty}P_{err}=0$
for the QB net of
Eq.(\ref{eq-qua-net}).
\end{claim}
\proof
Suppose $\rho\in \calp_{fac}$
and  $\vec{R}\in \calr(\rho)$.

We consider all
wiretap coding protocols that can be
described by the following QB net
\beq
\begin{array}{l}
\tr_{\rvr_{out}}
\\
\tr_{\rvr^n}
\\
\tr_{\rvr_\rvs}
\\
\tr_{\rvr_\rvt}
\end{array}
\sandb{
\entrymodifiers={+++[o][F-]}
\xymatrix@R=7pt{
\what{\rvs}&\cancel{\rvy^n}\ar[dl]\ar[l]&*{}
&*{}&\rvv^n\ar[dl]&\rvs\ar[l]
\\
\what{\rvt}_\rvy&*{}
&\scriptstyle\rvy^n,\rvz^n\ar[ul]_>>{\delta}\ar[dl]_>>{\delta}
&\cancel{\rvx^n}\ar[l]&*{}&*{}
\\
\what{\rvt}_\rvz&\cancel{\rvz^n}\ar[l]&*{}
&*{}&\rvu^n\ar[ul]\ar[uu]&\rvt\ar[l]
\\
*{}&\cancel{\rv{\calc}}\ar@/_1pc/[rruu]\ar[lu]\ar[luu]\ar[luuu]\ar@/_1pc/[rrru]\ar@/_2pc/[rrruuu]
&*{}&*{}&*{}&*{}
}
\\
\rvr_{out}\larrow \rvy^n,\rvz^n
\\
\rvr^n\larrow \rvy^n,\rvz^n,\rvv^n,\rvu^n
\\
\rvr_\rvs\larrow \rvs
\\
\rvr_\rvt\larrow \rvt
}
\sandb{\hc}
\;
\label{eq-qua-extra-net}
\eeq

We will use a codebook
$\calc=(\calc_\rvx,\calc_\rvv,\calc_\rvu)$
composed
of
 3 sub-codebooks. Codebook
$\calc_\rvx$ is
as an $N_\rvs N_\rvt\times n $ matrix
given by
$\calc_\rvx = \{x^n(s,t)\}_{\forall s,t}= x^n(\cdot)$
where $x^n(s,t)\in S^n_\rvx$ for all $(s,t)
\in S_\rvs\times S_\rvt$.
Similarly,
$\calc_\rvv = \{v^n(s,t)\}_{\forall s,t}= v^n(\cdot)$
where $v^n(s,t)\in S^n_\rvv$ for all $(s,t)$
and $\calc_\rvu = \{u^n(t)\}_{\forall t}= u^n(\cdot)$
where $u^n(t)\in S^n_\rvu$ for all $t$.

We will use
the shorthand notations
$\what{M}= (\what{s},\what{t}_\rvy,\what{t}_\rvz)$
and
$m=(s,t)$. Let $f$ be the function
that maps
$f(m)=f(s,t) = (s,t,t)$
for all $s,t$.

We assign
to the QB net of Eq.(\ref{eq-qua-extra-net})
the following node transition matrices:

\beq
A(s)=\frac{1}{\sqrt{N_\rvs}}
\;,\;\;
A(t)=\frac{1}{\sqrt{N_\rvt}}
\;,
\eeq

\beq
A(u^n|t,\calc)=
\delta(u^n,u^n(t))
\;,
\eeq

\beq
A(v^n|s,u^n,\calc)=
\delta(v^n,v^n(s,t))
\;,
\eeq
where $v^n(s,t)= v^n(s,u^n(t))$,

\beq
A(x^n|v^n,u^n,\calc)=\delta(x^n, x^n(s,t))
\;,
\eeq
where $x^n(s,t)= x^n(v^n(s,t),u^n(t))$,

\beq
A(y^n,z^n|x^n)=\prod_j A(y_j,z_j|x_j)
\;,
\eeq

\beq
A(\calc)=
\prod_{s,t}
A(x^n(s,t),v^n(s,t),u^n(t))
=
\prod
_{s,t,j}
A(x_j(s,t),v_j(s,t),u_j(t))
\;,
\eeq
where
$A(v,u)=A_{\rvv|\rvu}(v|u)A_\rvu(u)$
and
$A(x,v,u)=
A_{\rvx|\rvv,\rvu}(x|v,u)
A(v,u)$,
and

\beq
A(\what{M}|
y^n,z^n,\calc)
=
\prod_{\mu=1}^5
\prod_{m_\mu\neq \what{m}_\mu}
\theta(R_\mu< \Gamma_\mu)
\;,
\eeq
where the quantities $m_\mu$,
$R_\mu$
and $\Gamma_\mu$
are defined in Appendix
\ref{ap-a}. Assume that we are given
the reservoir amplitudes:
$A(r_{out}|y^n,z^n)$ (we choose this
one to be an \emph{isometry}),
$A(r^n|y^n,z^n,v^n,u^n)=\prod_j A(r_j|y_j,z_j,v_j,u_j)$,
$A(r_\rvs|s)$ and
$A(r_\rvt|t)$. Assume that we are also given
an \emph{isometry} $A(y,z|x)$ (called
the channel amplitude for this problem).

The reservoir amplitude
$A(r_{out}|y^n,z^n)$ is
applied by the decoder person
and is designed to be an isometry.
In a moment, we shall explain the reason why
we need it to be an isometry.
The reservoir amplitude
$A(r^n|y^n,z^n,v^n,u^n)$
need not be an isometry. It
is out of the control of any human
agent. As we shall see,
as long as it factors into $n$ amplitudes,
our results (for the
measure of success in quantum communication)
are independent of its value.
Our results will
also be independent of
the reservoir
amplitudes
$A(r_\rvs|s)$ and
$A(r_\rvt|t)$
acting on the inputs.

Henceforth we will use
$\av{\calt}$ as shorthand for

\beq
\av{\calt}=
\bra{\what{M}}
\calt_{\rv{\what{M}}|\rvm}
\left(\ket{m_0}\bra{m_0'}\right)
\ket{\what{M}'}
\;,
\eeq
and $A_{r^n}$ for

\beq
A_{r^n} = A(r^n|y^n,z^n,v^n(m_0),u^n(t_0))
\;,
\eeq
where $m_0=(s_0,t_0)$.
One finds for the above QB net
with the given node transition matrices
that

\beq
\av{\calt}=
\sum_{r^n,r_{out}}
\sandb{\sum_{y^n,z^n}
\cale_\calc
\left\{
\begin{array}{l}
A(r_{out}|y^n,z^n)
\\
A_{r^n} A(y^n, z^n| x^n(m_0))
\prod_{\mu=1}^5
\prod_{m_\mu\neq \what{m}_\mu}
\theta(R_\mu< \Gamma_\mu)
\end{array}
\right\}
}
\sandb{\hc\\
m_0\rarrow m_0'\\
\what{M}\rarrow \what{M}'
}
\;,
\eeq
where $\cale_\calc=
\sum_\calc A(\calc)$.

We immediately perform the sum
over $r_{out}$.
Since
$A(r_{out}|y^n,z^n)$
is an isometry,
upon summing over $r_{out}$,
the random variables $y^n$ and $z^n$
become classical.
Making
$y^n$ and $z^n$ classical
random variables
is the raison d'\^{e}tre for
the isometry $A(r_{out}|y^n,z^n)$.
Thus, we get

\beq
\av{\calt}=
\sum_{r^n,y^n,z^n}
\sandb{
\cale_\calc
A_{r^n}
A(y^n, z^n| x^n(m_0))
\prod_{\mu=1}^5
\prod_{m_\mu\neq \what{m}_\mu}
\theta(R_\mu< \Gamma_\mu)
}
\sandb{\hc\\
m_0\rarrow m_0'\\
\what{M}\rarrow \what{M}'
}
\;.
\label{eq-qua-psuc-pre-int_k}
\eeq

Let
\beq
\oint_k
=
\prod_{\mu=1}^5
\prod_{m_\mu\neq \what{m}_\mu}
\left\{
\int_{-\infty}^{+\infty}
\frac{dk_\mu(m_\mu)}{2\pi i}
\;\;
\frac{1}{(k_\mu(m_\mu)-i\eps)}
\right\}
\;,
\eeq
and

\beq
K_\mu = \sum_{m_\mu\neq \what{m}_\mu}
k_\mu(m_\mu)
\;.
\eeq
We will also use the following shorthand
notation

\beq
\what{x} = x(\what{s},\what{t})
\;,\;\;
\what{v} = v(\what{s},\what{t})
\;,\;\;
\what{u}= u(\what{t})
\;,
\eeq

\beq
\xi=(r,y,z,x(\cdot), v(\cdot),u(\cdot))
\;,
\eeq

\beq
A_r = A(r|y,z,v(m_0),u(t_0))
\;,
\eeq
and

\beq
\what{A}_r=
A(r|y,z,\what{v},\what{u})
\;.
\eeq

Expressing the $\theta$
functions in Eq.(\ref{eq-qua-psuc-pre-int_k})
 as
integrals,
we get

\beq
\av{\calt}=
\sum_{r^n,y^n,z^n}
\sandb{
\oint_k
e^{-i\sum_\mu K_\mu R_\mu}
\sum_{x^n(\cdot), v^n(\cdot),u^n(\cdot)}
e^{
\sum_{\xi}
n\ptype{\;}
(\xi)
\ln Z(\xi)
}
}
\sandb{\hc\\
m_0\rarrow m_0'\\
\what{M}\rarrow \what{M}'\\
}
\;,
\label{eq-qua-psuc-pre-p-type-int}
\eeq
where

\beqa
 Z(\xi)&=&
A_r A(y,z|x(m_0))
\prod_{s,t}\{
{\scriptstyle A(x(s,t),v(s,t),u(t))}
\}
\prod_{\mu=1}^5
\prod_{m_\mu\neq \what{m}_\mu}
\left\{
\gamma_\mu^{i\frac{k_\mu(m_\mu)}{n}}
\right\}
\\
&=&
\left\{
\begin{array}{l}
A_r A(y,z|x(m_0))
\prod_{s,t}
\left\{
\begin{array}{c}A(x(s,t)|v(s,t),u(t))\end{array}
\right\}
\\
\prod_{s,t}
\left\{
\begin{array}{c}A(v(s,t),u(t))\end{array}
\right\}
\prod_{\mu=1}^5
\prod_{m_\mu\neq \what{m}_\mu}
\left\{
\gamma_\mu^{i\frac{k_\mu(m_\mu)}{n}}
\right\}
\end{array}\right.
\;,
\eeqa
where the quantities $\gamma_\mu$
are defined in Appendix \ref{ap-a}.

Next we express
the sum over $x^n(\cdot), v^n(\cdot),u^n(\cdot)$
in Eq.(\ref{eq-qua-psuc-pre-p-type-int})
as a p-type integral and
do the p-type integration
by the steepest descent method. We
get for the leading order term:

\beq
\av{\calt}=
\sum_{r^n,y^n,z^n}
\sandb{
\oint_k
e^{-i\sum_\mu K_\mu R_\mu}
e^{
\sum_{r,y,z}
n\ptype{\;}
(r,y,z)
\ln Z(r,y,z)
}
}
\sandb{\hc\\
m_0\rarrow m_0'\\
\what{M}\rarrow \what{M}'\\
}
\;,
\label{eq-qua-pre-delta}
\eeq
where

\beq
Z(r,y,z)=
\sum_{x(\cdot),v(\cdot),u(\cdot)}
Z(\xi)
\;.
\eeq
If we define

\beq
A(y,z|v(\cdot),u(\cdot),m_0)
=
\sum_{x(\cdot)}
A(y,z|x(m_0))
\prod_{s,t}
A(x(s,t)|v(s,t),u(t)
)
\;,
\eeq
then $Z(r,y,z)$ can be expressed as

\beq
Z(r,y,z)
=
\cale_{v(\cdot),u(\cdot)}
A_r
A(y,z|v(\cdot),u(\cdot),m_0)
\prod_{m_\mu\neq \what{m}_\mu}
\left\{
\gamma_\mu^{i\frac{k_\mu(m_\mu)}{n}}
\right\}
\;.
\label{eq-qua-z-exp}
\eeq

Define

\beq
Z_{mid}(r,y,z)=
\cale_{\what{v},\what{u}}
\what{A}_r
A(y,z|\what{v},\what{u})
\prod_\mu
\left\{
\Phi_\mu^{i\frac{K_\mu}{n}}
\right\}
\;,
\eeq
where

\beq
A(y,z|v,u)=
\sum_x A(y,z|x)A(x|v,u)
\;,
\eeq
and where
the quantities $\Phi_\mu$
are defined in Appendix $\ref{ap-a}$.

Define
the integration operators

\beq
1_{h,K}=
\prod_\mu\left\{
\int_{-\infty}^{+\infty}dK_\mu\;
\int_{-\infty}^{+\infty}\frac{dh_\mu}{2\pi}\;
e^{-iK_\mu h_\mu}
\right\}
\;
\eeq
and

\beq
1_{h>0,K}=
1_{h,K}\prod_\mu\theta(h_\mu>0)
\;.
\eeq

Note that 1 equals

\beqa
1 &=& \prod_\mu\left\{
\int_{-\infty}^{+\infty}dK_\mu\;
\delta(\sum_{m_\mu\neq \what{m}_\mu}
\left\{k_\mu(m_\mu)\right\}-K_\mu)
\right\}
\\
&=&
1_{h,K}
e^{i\sum_\mu h_\mu
\sum_{m_\mu\neq \what{m}_\mu}k_\mu(m_\mu)
}
\;.
\label{eq-qua-one-is}
\eeqa
Multiplying $\av{\calt}$
by 1 certainly doesn't change it.
Thus
the right hand sides of
Eqs.(\ref{eq-qua-pre-delta})
and (\ref{eq-qua-one-is})
can be
multiplied to get

\beq
\av{\calt}=
\sum_{r^n,y^n,z^n}
\sandb{
1_{h,K}
e^{-i\sum_\mu K_\mu R_\mu}
\oint_k
e^{i
\sum_{m_\mu\neq \what{m}_\mu}k_\mu(m_\mu)}
e^{
\sum_{r,y,z}
n\ptype{\;}
(r,y,z)
\ln Z(r,y,z)
}
}
\sandb{\hc\\
m_0\rarrow m_0'\\
\what{M}\rarrow \what{M}'\\
}
\;.
\label{eq-qua-pre-km-int}
\eeq

Next
we will assume that,
for all $m_\mu$,
when doing the contour
integration over $k_\mu(m_\mu)$
in Eq.(\ref{eq-qua-pre-km-int})
with $Z(r,y,z)$
given by Eq.(\ref{eq-qua-z-exp}),
the
$e^{n\ln Z(r,y,z)}$
can be evaluated at the value
$k_\mu(m_\mu)=i\eps\rarrow 0$ of the pole.
{\it Furthermore,
we will assume that only the terms
with $\what{M}=m_0$
give a non-vanishing
contribution to the integral.}
Symbolically, this means we will assume

\beqa
\lefteqn{
\oint_k
e^{i\sum_\mu
h_\mu\sum_{m_\mu\neq \what{m}_\mu}k_\mu(m_\mu)}
e^{
\sum_{r,y,z}
n\ptype{\;}
(r,y,z)
\ln Z(r,y,z)
}
=}
\nonumber
\\
&=&
\delta_{\what{M}}^{f(m_0)}
e^{
\sum_{r,y,z}
n\ptype{\;}
(r,y,z)
\ln Z_{mid}(r,y,z)
}
\oint_k
e^{i\sum_\mu h_\mu\sum_{m_\mu\neq \what{m}_\mu}k_\mu(m_\mu)}
\\
&=&
\delta_{\what{M}}^{f(m_0)}
e^{
\sum_{r,y,z}
n\ptype{\;}
(r,y,z)
\ln Z_{mid}(r,y,z)
}
\prod_\mu\left\{
\theta(h_\mu >0)
\right\}
\;.
\label{eq-qua-magic-contour-int}
\eeqa
Applying Eq.(\ref{eq-qua-magic-contour-int})
to Eq.(\ref{eq-qua-pre-km-int}) gives

\beq
\av{\calt}=
\delta_{\what{M}}^{f(m_0)}
\delta_{\what{M}'}^{f(m_0')}
\sum_{r^n,y^n,z^n}
\sandb{
1_{h>0,K}
e^{-i\sum_\mu K_\mu R_\mu}
e^{
\sum_{r,y,z}
n\ptype{\;}
(r,y,z)
\ln Z_{mid}(r,y,z)
}
}
\sandb{\hc
}
\label{eq-qua-pre-yn-zn-sum}
\;.
\eeq

Next we express
the sum over $r^n,y^n,z^n$
in Eq.(\ref{eq-qua-pre-yn-zn-sum})
as a p-type integral and
do the p-type integration
by the steepest descent method. We
get for the leading order term:

\beq
\av{\calt}=
\delta_{\what{M}}^{f(m_0)}
\delta_{\what{M}'}^{f(m_0')}
1_{h>0,K}
1^*_{h'>0,K'}
e^{-i\sum_\mu(K_\mu-K'_\mu)R_\mu}
e^{n\ln
Z_{fin}}
\;,
\label{eq-qua-pre-sum-inside-exp}
\eeq
where

\beqa
Z_{fin}&=&
\sum_{r,y,z}
\sandb{ Z_{mid}(r,y,z)}
\sandb{\hc\\K_\mu\rarrow K_\mu'}
\\
&=&
\sum_{r,y,z}
\cale_{\what{v},\what{u}}
\cale_{\what{v}',\what{u}'}
\sandb{
\what{A}_r A(y,z|\what{v},\what{u})
}
\left(
\begin{array}{c}
1
\\
+\frac{i}{n}
\sum_\mu
K_\mu \ln \Phi_\mu
\\
-\frac{i}{n}
\sum_\mu
K'_\mu \ln \Phi'_\mu
\end{array}
\right)
\sandb{\hc
\\
\what{v}\rarrow\what{v}'
\\
\what{u}\rarrow\what{u}'
}
\;,
\label{eq-qua-zfin-pre-diag}
\eeqa
where we approximated

\beq
\prod_\mu \Phi_\mu^{i\frac{K_\mu}{n}}
\approx
1 + \frac{i}{n}
\sum_\mu
K_\mu \ln \Phi_\mu
\;.
\eeq
Because $\rho\in \calp_{fac}$,
we may set
in the right
hand side of Eq.(\ref{eq-qua-zfin-pre-diag})

\beq
A(y,z|\what{v},\what{u})=
A(y|\what{v},\what{u})
A(z|\what{v},\what{u})
\;,
\eeq
where
$A(y|\what{v},\what{u})$ and
$A(z|\what{v},\what{u})$ are both isometries.
It is important to
notice that $\sum_\mu
K_\mu \ln \Phi_\mu$
is a sum of terms
each of which only depends
either on $y$ or $z$ {\it but
not both}.
Thus in
Eq.(\ref{eq-qua-zfin-pre-diag}),
either the $y$ sum or the $z$ sum
(or both) kills
all terms except those with
 $\what{u}=\what{u}'$
and $\what{v}=\what{v}'$.
Hence

\beqa
Z_{fin}
&=&
\sum_{r,y,z,\what{v},\what{u}}
|\what{A}_r
A(y,z|\what{v},\what{u})
A(\what{v},\what{u})|^2
\left(1
+\frac{i}{n}
\sum_\mu
(K_\mu - K'_\mu)\ln \Phi_\mu
\right)
\\
&=&
1+\frac{i}{n}
\sum_\mu
(K_\mu-K'_\mu)
E_{y,z,\what{v},\what{u}}
\ln \Phi_\mu
\;,
\eeqa

\beq
n\ln Z_{fin}\approx
i
\sum_\mu
(K_\mu-K'_\mu)
E_{y,z,\what{v},\what{u}}
\ln \Phi_\mu
\;.
\label{eq-qua-short-sum-inside-exp}
\eeq
Combining Eqs.(\ref{eq-qua-pre-sum-inside-exp}) and
(\ref{eq-qua-short-sum-inside-exp})
yields

\beqa
\av{\calt}&=&
\delta_{\what{M}}^{f(m_0)}
\delta_{\what{M}'}^{f(m_0')}
\left|
1_{h>0,K}
e^{-i\sum_\mu K_\mu R_\mu}
e^{i\sum_\mu K_\mu
E_{y,z,\what{v},\what{u}}
\ln \Phi_\mu
}
\right|^2
\\
&=&
\delta_{\what{M}}^{f(m_0)}
\delta_{\what{M}'}^{f(m_0')}
\prod_\mu
\theta(R_\mu<S_\mu)
\\
&=&
\delta_{\what{M}}^{f(m_0)}
\delta_{\what{M}'}^{f(m_0')}
\theta(\vec{R}\in \calr(\rho))
\;,
\eeqa
where the quantities $S_\mu$
are defined in Appendix \ref{ap-a}.
\qed

\subsection{Capacity}
All statements\footnote{
Except for Claim \ref{claim-markov-diff}.
In the quantum case,
Claim
\ref{claim-markov-diff}
is true with $H() \rarrow S()$ if
one adds to the claim the additional
assumption that $\rve$
is a classical random variable.
For if $\rve$ is a classical
random variable
at the middle of a Markov chain, then we
can assert that
$S(\rva:\rvb|\rve)=0$.
}
made in Section \ref{sec-cla-cap}
for the channel
capacity for classical
wiretap coding
are also valid
for a quantum wiretap coding,
as long as (1)
classical informations
$H_P$ are replaced
by quantum informations $S_\rho$
throughout, and (2)
the symbols
$\calp_{gen}$, $\calp_{fac}$ and
$\calr(\calp)$
are
defined as
they were defined for quantum
instead of
classical wiretap coding.

\appendix
\section{Appendix: Table of Quantities Used In\\
Both Classical and Quantum Wiretap Coding}
\label{ap-a}

\beq
\begin{array}{|l|l|l|l|l|l|l|}
\hline
\mu & m_\mu& R_\mu&\Gamma_\mu & \gamma_\mu & \Phi_\mu & H_\mu
\\ \hline\hline
1&(s,t_\rvy,t_\rvz) & R_\rve&
\Gamma_{2y}-\Gamma_{2z}&
\frac{\gamma_{2y}}{\gamma_{2z}}&\frac{\Phi_{2y}}{\Phi_{2z}}
&H(\rvy:\rvv|\rvu)-H(\rvz:\rvv|\rvu)
\\ \hline
2&(s,t_\rvy) & R_\rvs + R_\rvt&
\Gamma_{2y}+\Gamma_{1y} &
\gamma_{2y}\gamma_{1y}& \Phi_{2y}\Phi_{1y}
&H(\rvy:\rvv|\rvu)+H(\rvy:\rvu)
\\ \hline
3&(s,t_\rvz) &R_\rvs + R_\rvt&
\Gamma_{2y}+\Gamma_{1z} &
\gamma_{2y}\gamma_{1z}&\Phi_{2y}\Phi_{1z}
&H(\rvy:\rvv|\rvu)+H(\rvz:\rvu)
\\ \hline
4& t_\rvy & R_\rvt& \Gamma_{1y} &
 \gamma_{1y}&\Phi_{1y}
 &H(\rvy:\rvu)
\\ \hline
5&t_\rvz &R_\rvt &\Gamma_{1z} &
\gamma_{1z}&\Phi_{1z}&H(\rvz:\rvu)
\\
\hline
\end{array}
\;\label{eq-table-params}
\eeq

The last
column of the table of
Eq.(\ref{eq-table-params})
is for the classical case.
The $H$'s are evaluated
at some $P\in \calr(\calp_{gen})$.
In the quantum case,
$H_\mu$ should be replaced by $S_\mu$,
and
all $H()$'s should be replaced
by $S()$'s. The
$S()$'s are evaluated at
some $\rho\in \calr(\calp_{gen})$.

In the classical case, one defines
\beq
\Gamma_{1y}
=
\frac{1}{n}
\ln
\frac{P(y^n:u^n(\what{t}_\rvy))}
{P(y^n:u^n(t_\rvy))}
\yz
\;,
\label{eq-appendix-cla-begin}
\eeq

\beq
\gamma_{1y}
=
\frac{P(y:u(\what{t}_\rvy))}
{P(y:u(t_\rvy))}
\yz
\;,
\eeq

\beq
\Phi_{1y}
=
P(y:u(\what{t}))
\yz
\;,
\eeq

\beq
\Gamma_{2y}
=
\frac{1}{n}
\ln
\frac{P(y^n:v^n(\what{s},\what{t}_\rvy)|u^n(\what{t}_\rvy))}
{P(y^n:v^n(s,t_\rvy)|u^n(t_\rvy))}
\yz
\;,
\eeq

\beq
\gamma_{2y}
=
\frac{P(y:v(\what{s},\what{t}_\rvy)|u(\what{t}_\rvy))}
{P(y:v(s,t_\rvy)|u(t_\rvy))}
\yz
\;
\eeq
and

\beq
\Phi_{2y}
=
P(y:v(\what{s},\what{t})|u(\what{t}))
\yz
\;.
\eeq

Note that
$\Gamma_\mu$
and $\gamma_\mu$
are related by

\beq
\Gamma_\mu
=
\sum_{y\in S_\rvy}
\sum_{z\in S_\rvz}
\sum_{v(\cdot)\in S^{N_\rvs N_\rvt}_\rvv}
\sum_{u(\cdot)\in S^{N_\rvt}_\rvu}
\ptype{\;}(y,z,v(\cdot),u(\cdot))
\ln(\gamma_\mu)
\;.
\label{eq-appendix-cla-end}
\eeq

Note that $\Phi_\mu$ is
evaluated at $\what{t}_\rvy=\what{t}_\rvz=\what{t}$.

In the quantum case,
Eqs.(\ref{eq-appendix-cla-begin})
to (\ref{eq-appendix-cla-end})
must be changed as follows.

In the quantum case,
$\rho_{\rvy,\rvz,\rvv,\rvu}$ is
given. From it,
one can define
$P_{\rvy,\rvz,\rvv,\rvu}=|A_{\rvy,\rvz,\rvv,\rvu}|^2$.
Suppose
$\rvx,\rvy,\rve$
are three distinct
random variables
from the set $\{\rvy,\rvz,\rvv,\rvu\}$.
Define the function
$\phi:\RR^{>0}\rarrow \RR$ by

\beq
\phi(r)=r^r
\;.
\eeq
In Eqs.(\ref{eq-appendix-cla-begin})
to (\ref{eq-appendix-cla-end}),
replace any
$\frac{1}{n}\ln P(x^n:y^n|e^n)$
by\footnote{br=bridge}

\beq
\frac{1}{n}
\ln P^{br}(x^n:y^n|e^n)
=
\sum_{x,y,e}
\ptype{x^n,y^n,e^n}(x,y,e)
\left\{
\begin{array}{l}
\frac{1}{P(x,y,e)}\ln\phi(\lam_{x,y,e}
(\rho_{\rvx,\rvy,\rve}))
\\
+
\frac{1}{P(e)}\ln\phi(\lam_{e}
(\rho_{\rve}))
\\
-
\frac{1}{P(x,e)}\ln\phi(\lam_{x,e}
(\rho_{\rvx,\rve}))
\\
-
\frac{1}{P(y,e)}\ln\phi(\lam_{y,e}
(\rho_{\rvy,\rve}))
\end{array}
\right\}
\;,
\label{eq-br-cmi-n}
\eeq
and any
$\ln P(x:y|e)$ by

\beq
\ln P^{br}(x:y|e)=
\mbox{expression inside
curly brackets in Eq.(\ref{eq-br-cmi-n})}
\;.
\eeq
Also replace any
$\frac{1}{n}\ln P(x^n:y^n)$
by

\beq
\frac{1}{n}
\ln P^{br}(x^n:y^n)
=
\sum_{x,y}
\ptype{x^n,y^n}(x,y)
\left\{
\begin{array}{l}
\frac{1}{P(x,y)}\ln\phi(\lam_{x,y}
(\rho_{\rvx,\rvy}))
\\
-
\frac{1}{P(x)}\ln\phi(\lam_{x}
(\rho_{\rvx}))
\\
-
\frac{1}{P(y)}\ln\phi(\lam_{y}
(\rho_{\rvy}))
\end{array}
\right\}
\;,
\label{eq-br-mi-n}
\eeq
and any
$\ln P(x:y)$ by

\beq
\ln P^{br}(x:y)=
\mbox{expression inside
curly brackets in Eq.(\ref{eq-br-mi-n})}
\;.
\eeq

{\it Note that the bridge
functions
$P^{br}$
depend on both
$\rho_{\rvx,\rvy,\rve}$
and $P_{\rvx,\rvy,\rve}$.
Note also that the bridge functions
equal the classical functions they
are replacing
when $\rho_{\rvx,\rvy,\rve}$
is diagonal (the classical case)}.


\begin{thebibliography}{99}
\bibitem{CovTh}
Thomas M. Cover, Joy A. Thomas,
{\it Elements of Information Theory}
(Wiley-Interscience, 1991)

\bibitem{Wilde}Mark M. Wilde,
From Classical to Quantum Shannon Theory,
arXiv:1106.1445

\bibitem{ElG}Abbas El Gamal, Young-Han Kim,
 Lecture Notes on Network Information Theory,
 arXiv:1001.3404

 \bibitem{Wy}A.D. Wyner,
 ``The wire-tap channel",
 Bell Syst. Tech. J.,
 vol.54, no. 8, pp. 1355-1387, 1975

\bibitem{CK}
I. Csisz\'{a}r, J. K\"{o}rner,
``Broadcast Channels
with Confidential Messages",
IEEE, Trans. in Info. Th.,
Vol 24, No.3, May 1978.

\bibitem{Tuc-mixology}
R.R. Tucci,
``An Introduction to
Quantum Bayesian Networks for
Mixed States",
arXiv:1204.1550

\bibitem{Tuc-redoing-classics}
R.R. Tucci,
``Shannon Information Theory
Without Shedding Tears
Over Delta \& Epsilon Proofs
or Typical Sequences",
arXiv:1208.2737

\bibitem{Tuc-inequalities}
R.R. Tucci,
``Some Quantum Information Inequalities
from a Quantum Bayesian Networks Perspective",
arXiv:1208.1503


\end{thebibliography}
\end{document}